\documentclass[twocolumn]{aastex63}

\usepackage{bm}
\newcommand{\vect}[1]{$\bm{#1}$}

\newcommand{\teff}{$T_{\mathrm{eff}}$}
\newcommand{\logg}{log~$g$}
\newcommand{\feh}{[Fe/H]}
\newcommand{\kms}{km~s$^{-1}$}

\newcommand{\afe}{[$\alpha$/Fe]}

\newcommand{\numstar}{98,736}
\newcommand{\numspec}{120,571}
\newcommand{\nummp}{416}

\newcommand{\project}[1]{\textsl{#1}}
\newcommand{\gaia}{\project{Gaia}}
\newcommand{\hetdex}{HETDEX}
\newcommand{\gaiahet}{Gaia-HETDEX}

\newcommand{\rave}{RAVE}
\newcommand{\galah}{GALAH}
\newcommand{\ges}{Gaia-ESO}
\newcommand{\apogee}{APOGEE}

\newcommand{\lamost}{LAMOST}

\newcommand{\sdss}{SDSS}
\newcommand{\segue}{SEGUE}

\accepted{February 11, 2021}
\shorttitle{The Stars in HETDEX iDR2}
\shortauthors{Hawkins et al.}
\graphicspath{{./}{figures/}}

\begin{document}

\title{The Stars of the HETDEX Survey. I. Radial Velocities and Metal-Poor Stars from Low-Resolution Stellar Spectra}

\correspondingauthor{Keith Hawkins}
\email{keithhawkins@utexas.edu}

\author[0000-0002-1423-2174]{Keith Hawkins}
\affiliation{Department of Astronomy, The University of Texas at Austin, 2515 Speedway Boulevard, Austin, TX 78712, USA}

\author{Greg Zeimann}
\affiliation{Hobby Eberly Telescope, University of Texas, Austin, Austin, TX, 78712}

\author[0000-0002-3456-5929]{Chris Sneden}
\affiliation{Department of Astronomy, The University of Texas at Austin, 2515 Speedway Boulevard, Austin, TX 78712, USA}

\author{Erin Mentuch Cooper}
\affiliation{Department of Astronomy, The University of Texas at Austin, 2515 Speedway Boulevard, Austin, TX 78712, USA}

\author{Karl Gebhardt}
\affiliation{Department of Astronomy, The University of Texas at Austin, 2515 Speedway Boulevard, Austin, TX 78712, USA}

\author[0000-0003-1377-7145]{Howard E. Bond}  
\affiliation{Department of Astronomy \& Astrophysics, The Pennsylvania State University, University Park, PA 16802, USA} 
\affiliation{Space Telescope Science Institute, 3700 San Martin Drive, Baltimore, MD 21218, USA}

\author{Andreia Carrillo}
\affiliation{Department of Astronomy, The University of Texas at Austin, 2515 Speedway Boulevard, Austin, TX 78712, USA} 

\author{Caitlin M. Casey}
\affiliation{Department of Astronomy, The University of Texas at Austin, 2515 Speedway Boulevard, Austin, TX 78712, USA}

\author{Barbara G. Castanheira}
\affiliation{Department of Physics, Baylor University, One Bear Place \#97316 Waco, TX 76798-7316, USA} 
\affiliation{McDonald Observatory, The University of Texas at Austin, 2515 Speedway Boulevard, Austin, TX 78712, USA}

\author{Robin Ciardullo}
\affil{Department of Astronomy \& Astrophysics, The Pennsylvania State University, University Park, PA 16802, USA}
\affil{Institute for Gravitation and the Cosmos, The Pennsylvania State University, University Park, PA 16802, USA}

\author{Dustin Davis}
\affiliation{Department of Astronomy, The University of Texas at Austin, 2515 Speedway Boulevard, Austin, TX 78712, USA}

\author{Daniel J. Farrow}
\affiliation{Max-Planck-Institut f{\"u}r extraterrestrische Physik, Postfach 1312 Giessenbachstrasse, 85741 Garching, Germany}

\author[0000-0001-8519-1130]{Steven L. Finkelstein}
\affiliation{Department of Astronomy, The University of Texas at Austin, 2515 Speedway Boulevard, Austin, TX 78712, USA}

\author{Gary J. Hill}
\affiliation{Department of Astronomy, The University of Texas at Austin, 2515 Speedway Boulevard, Austin, TX 78712, USA} 
\affiliation{McDonald Observatory, The University of Texas at Austin, 2515 Speedway Boulevard, Austin, TX 78712, USA}

\author{Andreas Kelz}
\affiliation{Leibniz-Institut f{\"u}r Astrophysik Potsdam (AIP), An der Sternwarte 16, 
14482 Potsdam, Germany}

\author{Chenxu Liu}
\affiliation{Department of Astronomy, The University of Texas at Austin, 2515 Speedway Boulevard, Austin, TX 78712, USA}

\author{Matthew Shetrone}
\affiliation{UC Observatories UC Santa Cruz 1156 High Street Santa Cruz, CA 95064}

\author{Donald P. Schneider}
\affiliation{Department of Astronomy \& Astrophysics, The Pennsylvania State University, University Park, PA 16802, USA}
\affil{Institute for Gravitation and the Cosmos, The Pennsylvania State University, University Park, PA 16802, USA}

\author{Else Starkenburg}
\affiliation{Leibniz-Institut f{\"u}r Astrophysik Potsdam (AIP), An der Sternwarte 16, 
14482 Potsdam, Germany}
\affiliation{Kapteyn Astronomical Institute, University of Groningen, Landleven 12, 9747 AD Groningen, The Netherlands}

 \author[0000-0001-6516-7459]{Matthias Steinmetz}
\affiliation{Leibniz-Institut f{\"u}r Astrophysik Potsdam (AIP), An der Sternwarte 16, 
14482 Potsdam, Germany}

\author{Craig Wheeler}
\affiliation{Department of Astronomy, The University of Texas at Austin, 2515 Speedway Boulevard, Austin, TX 78712, USA}

\collaboration{30}{(The HETDEX Collaboration)}

\begin{abstract}
The Hobby-Eberly Telescope Dark Energy Experiment (\hetdex) is an unbiased, massively multiplexed spectroscopic survey, designed to measure the expansion history of the universe through low-resolution ($R\sim750$) spectra of Lyman-Alpha Emitters. In its search for these galaxies, \hetdex\ will also observe a few 10$^{5}$ stars. In this paper, we present the first stellar value-added catalog within the internal second data release of the \hetdex\ Survey (HDR2). The new catalog contains \numspec\ low-resolution  spectra for \numstar\ unique stars between $10 < G < 22$ spread across the \hetdex\ footprint at relatively high ($b\sim60^\circ$) Galactic latitudes. With these spectra, we measure radial velocities (RVs) for $\sim$42,000 unique FGK-type stars in the catalog and show that the \hetdex\ spectra are sufficient to constrain these RVs with a 1$\sigma$ precision of 28.0~\kms\ and bias of 3.5~\kms\ with respect to the \lamost\ surveys and 1$\sigma$ precision of 27.5~\kms\ and bias of 14.0~\kms\ compared to the \segue\ survey. Since these RVs are for faint ($G\geq16$) stars, they will be complementary to \gaia. Using t-Distributed Stochastic Neighbor Embedding (t-SNE), we also demonstrate that the \hetdex\ spectra can be used to determine a star's \teff, and \logg\ and its \feh. With the t-SNE projection of the FGK-type stars with \hetdex\ spectra we also identify \nummp\ new candidate metal-poor (\feh\ $< -1$~dex) stars for future study. These encouraging results illustrate the utility of future low-resolution stellar spectroscopic surveys. 
\end{abstract}

\keywords{catalogs -- stars: general	--- surveys -- methods:statistical}

\section{Introduction} \label{sec:intro}
One of the primary goals within Galactic astronomy is to constrain the nature of the physical processes that govern how the Milky Way formed, assembled, and evolved from its birth to today. The key observational data that constrains theories of Milky Way formation include the position, velocities, and stellar atmospheric parameters for millions of stars across the Galaxy. This is what has given rise to missions such as \gaia, which have enabled the detailed study of the positions and velocities for $\sim$1.7 billion stars through precise astrometric data \citep{Gaiasummary2018}, and has also been a key driver for the creation of large multi-object stellar spectroscopic surveys. Stellar spectra, whether at low- or high-resolution\footnote{Resolving power is defined as $R = \lambda/\Delta\lambda$.}, are particularly useful because they enable not only measurement of the radial velocity (RV) of stars, giving us 3-D velocity information when combined with \gaia, but also they contain critical information, including the effective temperature (\teff), surface gravity (\logg), metallicity (\feh), and chemical abundance ratios ([X/Fe]), for large samples of stars. 

Large spectroscopic surveys have been instrumental in uncovering the history and nature of our Milky Way Galaxy and beyond. These projects have obtained data, which range from low-to-high resolution. For example, while the primary goal of the original the Sloan Digital Sky Survey \citep[\sdss,][]{York2000} was to map galaxies and quasars across the universe, the survey also obtained low-resolution (R$\sim$2000) spectra for $\sim10^5$ stars through the Sloan Extension for Galactic Understanding and Exploration \cite[SEGUE,][]{Yanny2009}. Similar to SEGUE, the most recent data release (DR5) from the Large Sky Area Multi-Object Fibre Spectroscopic Telescope survey \citep[\lamost\ DR5,][]{Luo2015, Xiang2017} contains low-resolution (R$\sim$1800) spectra for over 8 million stars. At moderate resolution (R$\sim$7500), the magnitude-limited Radial VElocity Experiment \citep[RAVE,][]{Steinmetz2006,Steinmetz2020}, whose aim was to explore the structure and nature of the Milky Way, observed more than half a million stars observed in the southern hemisphere and derive not only RV informaiton but also stellar atmospheric parameters for their sample. 

Moving towards higher resolution the GALactic Archaeology with HERMES \citep[\galah,][]{De_silva2015, Buder2018} survey has obtained optical spectra of 342,682 stars at R$\sim$28,000, the \sdss-IV Apache Point Observatory Galactic Evolution Experiment \citep[\apogee,][J\"{o}nsson et al.\ in prep.]{Majewski2017, Ahumada2019} survey has collected infrared (H-band) spectra for upwards of 400,000 stars at R$\sim$21,000, and the \ges\ survey \citep[][]{Gilmore2012} aims to collect up to 100,000 spectra at R$\sim$47,000. As one moves toward higher resolution, sample sizes decrease because of the tradeoff that exists between the resolution of the spectrograph and the exposure time required to obtain high signal-to-noise ratio (SNR) data required for the scientific aims of the surveys. 

With high resolution, it is possible to derive both the detailed stellar atmospheric parameters (\teff, \logg, \feh) and the chemical abundance ratios, denoted as [X/Fe], for many elements. This is the case for surveys like \apogee, \galah, and \ges. However, for lower resolution spectroscopic surveys (e.g., \sdss, \lamost), it becomes significantly more challenging to recover the stellar parameters and detailed chemical fingerprint. In fact, until recently,  many low-resolution surveys only were able to determine the \teff, \logg, \feh, and, in some cases, \afe\footnote{In this case, \afe\ represents the average abundance ratio of [X/Fe] the $alpha$ elements (i.e. X = [Mg, Si, Ca, ...]).}  \citep[e.g.,][]{Lee2008, Lee2015, Boeche2018}. Despite this limitation, \cite{Ting2017b} demonstrated that it may be possible to measure not only the stellar atmospheric parameters but also several key individual elements down to R$\gtrapprox$700 with precisions at the $\sim$0.15~dex level. While, at first, the results from these authors seem surprising, \cite{Xiang2019} illustrated the power of new machine learning techniques and derived the stellar parameters and 16 [X/Fe] atmospheric abundance ratios from the low-resolution (R$\sim$1800) LAMOST spectra. Additionally, other works have shown that it is possible to uncover metal-poor stars from low-resolution (R$<$ 1000) spectra \citep[e.g.][]{Aguado2019} and even from photometry \citep[e.g.][]{Starkenburg2017, Casagrande2019, Lucey2019}. 

These new low-resolution techniques offer many opportunities including discovering and characterizing relatively rare, yet scientifically important, metal-poor stars. Metal-poor stars, defined as those with \feh~$<$~$-1.0$~dex, are important because they offer insights into the early Galaxy as well as constrain the nucleosynthesis of the elements \cite[e.g.][and references therein]{Bromm2004, Beers2005, Karlsson2013, Yong2013, Roederer2014, Frebel2015, Sakari2019}. These stars are pretty rare yet their chemical properties are highly sought after. Hunting metal-poor star, especially those at the lowest metallicities is still an critical task. Several low- to medium-resolution surveys have been instrumental in finding and characterizing these metal-poor stars \citep[e.g.][]{Aoki2008, Ruchti2010, Xu2013, Matijevic2017, Li2018}.  

In this context, the Hobby-Eberly Telescope Dark Energy Experiment \citep[\hetdex,][Gebhardt et al., in prep.]{Hill2008, Hill2016} is a low-resolution (R$\sim$750) {\it blind} spectroscopic survey whose primarily goal is to map the locations and redshifts of high-$z$ Lyman-$\alpha$ emitting galaxies to constrain the nature of dark energy. This survey will cover a total of $\sim$450 sq. deg. However, much like the case for \sdss, which was designed for cosmological investigations but eventually observed many stars, \hetdex\ will obtain spectra for large numbers of stars in its search for Lyman-$\alpha$ emitters. The unbiased nature of \hetdex\ (i.e., no pre-selection of objects is done), makes it extremely powerful, when compared to other large stellar spectroscopic surveys, to answer questions about the properties of the Milky Way. Therefore, we explore the use of these low-resolution \hetdex\ data for stellar astrophysics applications, such as deriving radial velocities for faint stars in the \gaia\ catalogue and finding rare, but scientifically critical, metal-poor stars using machine learning. This is the first time these techniques have been applied to this dataset. 

This paper is structured as follows: Section~\ref{sec:data} details the spectroscopic data collected from the \hetdex\ survey and presents the \hetdex-\gaia\ value added catalogue. We also describe the overlap data set used to validate our methods. In section~\ref{sec:methods} we describe the methods used to constrain the RVs for the FGK-type stars found in the \hetdex\ survey and the t-distributed Stochastic Neighbor Embedding (tSNE) machine learning method that we used to uncover candidate metal-poor stars. Section~\ref{sec:results} presents the results of the RV and tSNE analysis and discusses the role \hetdex\ may play in constraining the nature of the Galaxy. Finally, in section~\ref{sec:summary} we summarize our results and describes several future possible investigations with the data presented here.

\section{Data} \label{sec:data}
\subsection{The HETDEX Survey} \label{subsec:HETDEX} 
\begin{figure*}
	 \includegraphics[width=2\columnwidth]{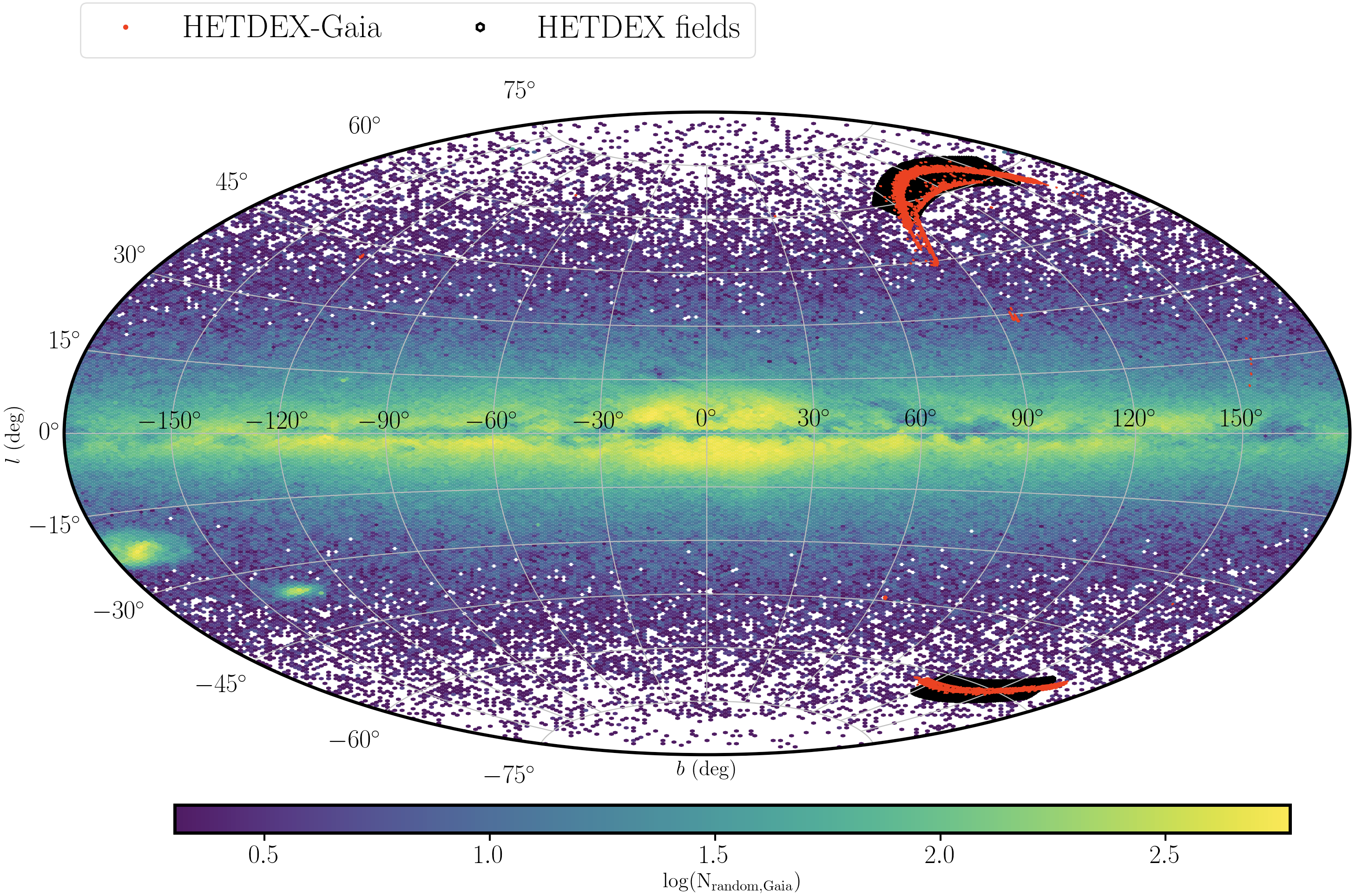}
	\caption{An Aitoff projection of the Galactic coordinates for the \hetdex\ fields (black pentagons) and stars observed in this data release (red circles). For reference, we also show positions for 2 million randomly selected stars from the \gaia\ mission. While most of the stars lie in the main \hetdex\ footprint (at high Galactic latitude), there are some that can also be found in early calibration and extension fields.} 
	\label{fig:sky}
\end{figure*}

The \hetdex\ survey, which aims to constrain the nature of dark energy by observing Lyman-$\alpha$ emitters in the redshift interval 1.9~$\leq z \leq$~3.5, is currently being carried out on the Hobby-Eberly Telescope (HET; \citealt{Ramsey1994}), which is an 11-m telescope with a segmented primary mirror, recently upgraded with a 22 arcmin field of view and 10 m pupil \citep{Hill2018_wfu}. The upgraded HET includes a key new instrument called the Visible Integral-field Replicable Unit Spectrograph \citep[VIRUS,][]{Hill14, Hill2018_virus}, which is a massively replicated, fiber-fed integral field spectrograph designed specifically to preform unbiased spectroscopic surveys. The entire VIRUS instrument consists of up to 156 integral field low-resolution (resolving power, $R\sim$750) spectrographs, arrayed in 78 pairs. Each of the 78 VIRUS units is fed by 448 $1\farcs5$ fibers. Each of the fibers are  fed to individual spectrographs (two per IFU) and dispersed onto CCD detectors at a mostly equal spacing on the detector.
The spectrographs have a fixed spectral window in the optical with 3500 $< \lambda <$ 5500~\AA\ with 2~\AA\ pixels. This represents a pixel size of $\sim$130~\kms. At full deployment\footnote{The VIRUS units have been deployed on the HET one-by-one, beginning in 2015. As of writing this paper, 74 of the 78 units are currently active. The data from this paper were collected over the period January 2017 -- June 2020 with between 16 and 71 units deployed.}, VIRUS will be able to capture 34,944 spectra for each exposure. The instrumentation for HETDEX is described in Hill et al., 2021, in preparation. The \hetdex\ survey, which is currently underway with the VIRUS instrument, will be an {\it blind} emission line survey of an area on the sky comprising of $\sim 450$ sq. deg with a 1/4.5 fill factor. The current release of \hetdex\ contains data from $\sim$29.85 sq. deg, or $\sim$30\% of the expected final \hetdex\ dataset. Figure~\ref{fig:sky}, shows the location of the main \hetdex\ survey footprint (black) and a density map of 2 million randomly selected stars from the \gaia\ mission. The background collection of stars from \gaia\ is to provide a Galactic context to the \hetdex\ fields. Also shown is the Galactic coordinates of the stars (section~\ref{subsec:GaiaHETDEX}) in this data release of the \hetdex\ survey (red points). 

Each \hetdex\ pointing consists of a set of 3-dithered 6~min exposures  (done in order to fill the sky gaps between fibers; Gebhardt et al., 2021 in preparation). The \hetdex\ fields are mostly at high Galactic latitude, with a median absolute Galactic latitude of $\sim$60~deg (see Figure~\ref{fig:sky}). In \hetdex's quest to survey $\sim10^6$  Lyman-$\alpha$ emitters, 
it will also capture spectra for many ($\sim10^{5}$) stars. It is these stars that are the subject of this work. We make use of the second internal data release for the \hetdex\ survey, which we will refer to as iHDR2.

\subsection{The HETDEX-Gaia Value Added Catalog} \label{subsec:GaiaHETDEX}
The primary goal of this work is to build a catalouge of stellar spectra from observed within the \hetdex\ survey and use those spectra to (1) measure RVs for \gaia\ sources and (2) find metal-poor candidate stars. To achieve these aims we must first obtain a list of point sources (i.e., stars) in the \hetdex\ survey. This task is accomplished by cross matching each centroid of each fiber pointing that exists in iHDR2, which includes $\sim$200~million spectra, against the $\sim$1.7~billion point sources in the \gaia~DR2 catalogue \citep{Gaiasummary2018}. In order to determine the stars in common between \hetdex\ and \gaia, we  cross-match each fiber center in iHDR2 with \gaia~DR2 using a generous 3\arcsec\ search radius in RA and DEC\null. This relatively large cross-match radius compared to the $1\farcs 5$ fiber diameter of VIRUS ensures that we capture the stars that overlap between \hetdex\ and \gaia~DR2.  The result of this cross match is referred to as the \hetdex-\gaia\ value added catalogue. We note that distant galaxies can also appear as point sources and therefore we expect some level of contamination from galaxies in our sample.  

\begin{table*}
\caption{The \hetdex-\gaia\ Value Added Catalogue}
\label{tab:HGVAC}
\begin{tabular}{rrrrrrrrrr}
\hline\hline
Gaia Source ID & RA & DEC & $G$ & $BP-RP$ & SNR$_{\mathrm{HETDEX}}$ & RV & $\sigma$RV & Barycentric Correction & ... \\
 & (deg) & (deg) & (mag)& (mag) & (pixel$^{-1}$) & (\kms)  & (\kms)  & (\kms) & \\
 \hline
2543173457359409664 & 9.78194 & --0.00009 & 13.06 & 0.92 & 58.73 & --8.49 & 28.62 & 4.18 & .... \\
2543068690220755456 & 10.45149 & 0.00748 & 19.15 & 1.90 & 3.58 & --192.69 & 62.57 & 4.19 & .... \\
2543068690220758144 & 10.45224 & 0.01046 & 18.88 & 1.90 & 5.73 & 75.98 & 45.75 & 4.19 & .... \\
2543092398440275072 & 10.49391 & 0.05285 & 20.32 & 0.81 & 5.73 & --206.13 & 63.73 & 4.22 & .... \\
2543733211856949120 & 6.67326 & --0.02982 & 17.51 & 1.12 & 29.14 & --17.25 & 19.08 & --6.87 & .... \\
2543740010789142528 & 6.84129 & 0.01837 & 20.58 & 1.08 & 4.21 & --55.21 & 22.88 & --6.78 & .... \\
2543735509663099648 & 6.84963 & --0.00984 & 19.52 & 1.55 & 7.90 & --14.42 & 35.71 & --6.78 & .... \\
2543040137278030848 & 10.73567 & --0.04865 & 17.18 & 1.85 & 20.31 & --11.99 & 57.23 & --4.99 & .... \\
2543039381363775232 & 10.74644 & --0.05880 & 20.23 & 1.58 & 2.64 & 51.64 & 50.80 & --4.99 & .... \\
2543041030631244928 & 10.79130 & --0.03320 & 19.38 & 1.74 & 4.69 & --90.57 & 40.68 & --4.96 & .... \\
2543041408588483968 & 10.86433 & 0.00056 & 19.81 & 1.27 & 5.91 & --272.85 & 137.26 & 5.40 & .... \\
2543175415864489472 & 9.61114 & --0.01187 & 17.25 & 1.29 & 48.19 & --471.99 & 143.30 & 4.10 & .... \\
2543172009953849344 & 9.62519 & --0.04361 & 19.58 & 1.08 & 10.82 & --111.52 & 18.64 & 4.10 & .... \\
2543174728669728768 & 9.75072 & 0.05631 & 19.33 & 1.33 & 27.20 & 2.98 & 32.01 & 4.18 & .... \\
2543174385072345216 & 9.77910 & 0.02872 & 17.40 & 1.27 & 30.14 & --56.76 & 25.57 & 4.19 & .... \\
2543173457359409664 & 9.78194 & --0.00009 & 13.06 & 0.92 & 95.78 & 28.80 & 14.25 & 4.18 & .... \\
2543736579111320448 & 6.96626 & 0.01918 & 17.92 & 1.34 & 20.29 & --17.54 & 20.59 & --7.54 & .... \\
2543831343269590656 & 7.00635 & 0.06922 & 12.79 & 0.89 & 65.11 & --24.53 & 12.01 & --7.51 & .... \\
2543547463111076864 & 7.06080 & --0.05459 & 14.78 & 0.90 & 68.92 & --70.10 & 11.09 & --7.51 & .... \\
2543642295989347840 & 7.08448 & --0.00604 & 16.59 & 1.42 & 35.53 & 25.00 & 24.06 & --7.49 & .... \\
...&...&...&...&...&...&...&...& ... &...\\
\hline\hline
\end{tabular}
\raggedright

NOTE: This is a subsample of the full \hetdex-\gaia\ value added catalogue. The \gaia\ DR2 source identifier of each star is given in column~1 with equatorial coordinates (in J2015.5) in columns~2 and 3. The \gaia\ $G$ band magnitude and the $BP-RP$ color for each star are listed in columns 4 and 5, respectively. The average signal to noise ratio (SNR) per pixel estimated using the \hetdex\ spectra is given in column~6. The RV and its uncertainty measured from cross-correlating the \hetdex\ spectra against a template spectrum are listed in columns~7 and 8, respectively. The heliocentric correction, used to correct the measured RV into the rest frame of the Sun, can be found in column 9. This is only a subsample of the full table, which will be provided online, illustrate the the catalogue's format. 
\end{table*}

\begin{figure}
	 \includegraphics[width=1\columnwidth]{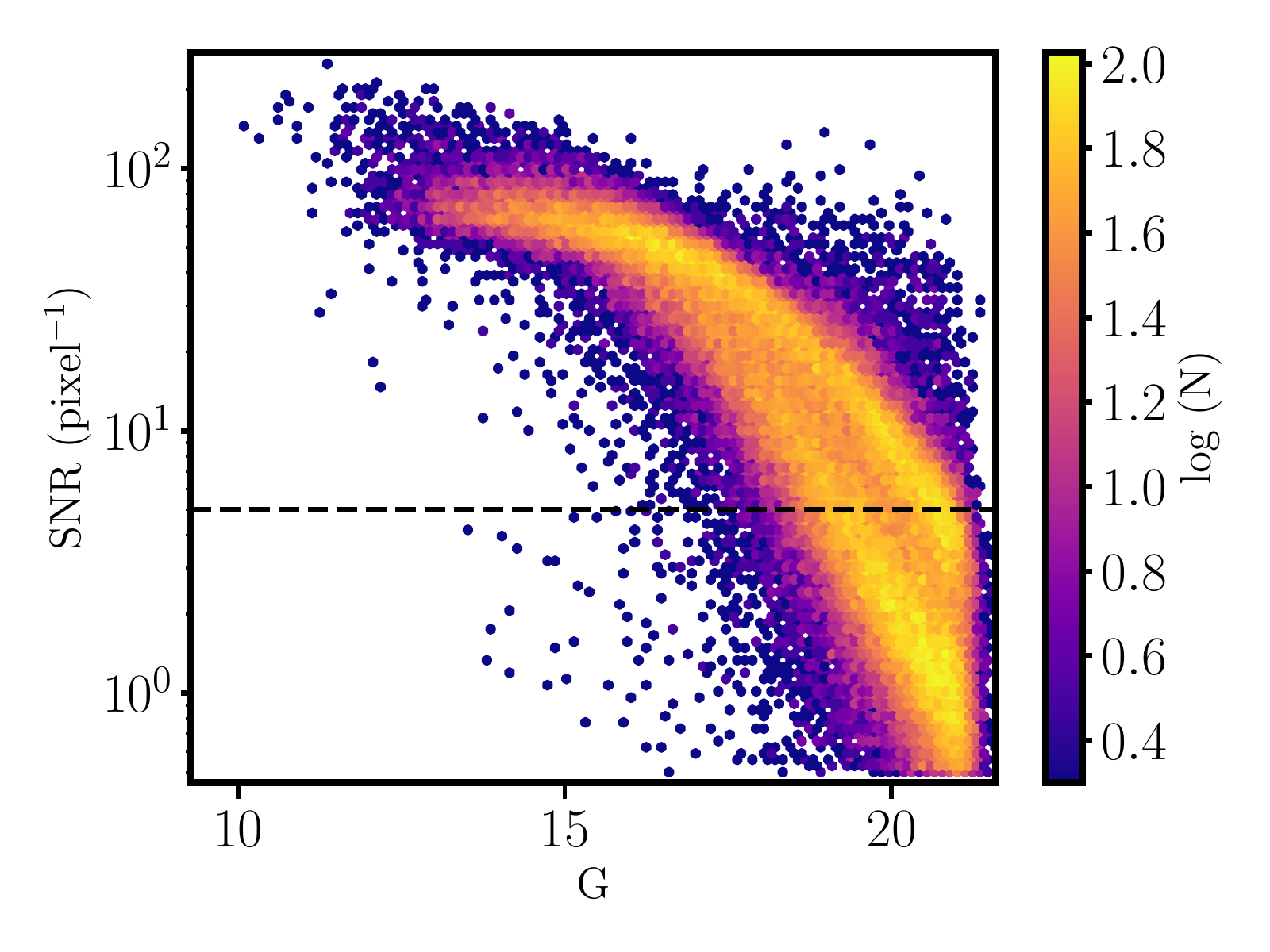}
	\caption{The median SNR per pixel for each star of the \numstar\ stars in the \hetdex-\gaia\ catalogue  as a function of their \gaia\ $G$ band magnitude. The black dashed line represents the SNR = 5~pixel$^{-1}$; we generally focus on the $\sim$47,000 sources above this line. } 
	\label{fig:SNR}
\end{figure}

The final \hetdex-\gaia\ value added catalogue contains \numstar\ unique point sources in the \gaia~DR2 catalogue with spectral data from \hetdex.  Table~\ref{tab:HGVAC} is subsample of the observational and spectroscopic properties of this catalogue. The full catalogue will be available online. With the cross-match in hand, we re-extracted the \hetdex\ spectra at the \gaia~DR2 positions accounting for proper motion on the date of the \hetdex\ observation. If a \gaia~DR2 source was observed on multiple nights, multiple independent extractions we preformed. We used an optimal method \citep{Horne1986} for extraction that employs a Moffat spatial profile \citep{Moffat1969} set to the seeing conditions of the observation as well as a maximum aperture of 3\arcsec.

\begin{figure}
	 \includegraphics[width=1\columnwidth]{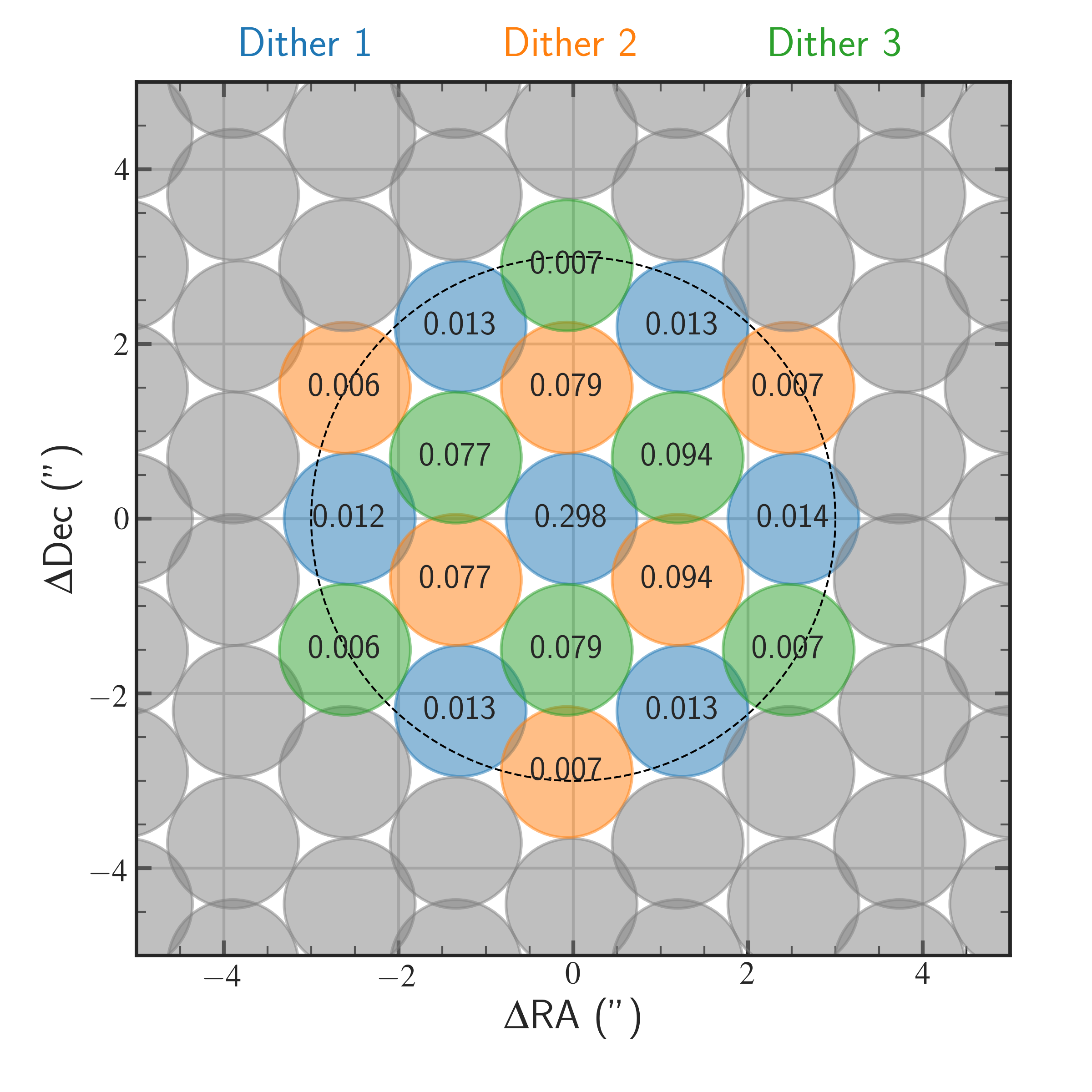}
	\caption{The fiber layout for an example source in an observation with 1.8” seeing. The source being at $\Delta$RA = $\Delta$DEC = 0. A HETDEX observation includes three dithers shown in blue, orange, and green.  The fiber diameter is 1.5” and in this example since the closest fiber from dither 1 is nearly centered on the source for extraction, it contains roughly 30\% of the total light from the Moffat spatial profile.  We show the 3” extraction aperture as a dashed black line and the aperture includes all fibers that overlap that radius.  We show the fiber coverage for each of the extracted fibers and these coverage values serve as the weights in our optimal extraction.  In total 94\% of the light is captured in the highlighted fibers.  The fibers outside of the extraction aperture are all grey.}
	\label{fig:fibers}
\end{figure}

To perform extractions for any individual source we collect all fibers within 7 arcseconds.  We use the seeing conditions of the observation to construct a Moffat spatial profile (Moffat 1969) and we integrate that profile over the 1.5 arcsecond diameter fibers.  The integral of the profile over the fiber is called the covering fraction.  This covering fraction depends on the distance of the source from the fiber center, the seeing conditions, and the wavelength as the HET does not correct for differential atmospheric refraction so the position of a source with respect to the fiber location changes as a function of wavelength.  We illustrate the fiber coverage fractions in Figure~\ref{fig:fibers}.  We use these fractions as weights in an optimal extraction \citep{Horne1986}. Each extracted spectrum includes the calibrated flux (in units of $\times10^{-17}$ ergs~cm$^{-2}$~s$^{-1}$), a 1$\sigma$ error array, and a fiber coverage array. The fiber coverage array is defined as the fraction of light that the \hetdex\ fibers collect for a given point source at a given wavelength. The sky background is estimated using the biweight of all fibers in the observation after the fibers have been flat-fielded using twilight observations.  We do not remove contamination from nearby sources. We exclude sources for which fiber coverage arrays contained only pixels with coverages less than 5\% (i.e., where $<$5\% of its light from the star made it into all extracted fibers). Detailed information on the re-extraction methods is included in Zeimann et al.\ (2021, in preparation). 

In addition, we also computed a `continuum normalized' version of the spectral fluxes, which are normalized using the boosted median continuum \citep[henceforth BMC, e.g.,][]{Rogers2010, Hawkins2014}. In this approach, the continuum to be divided is defined as a prescribed percentile of the flux in a predefined window. This is a common approach for low-resolution spectra for which the true continuum is extremely difficult to define due to significant line blending. For the purposes of the \hetdex\ spectra, we choose a window size of 50~\AA\ (i.e., 25 pixels) and a percentile of 98\%. Additionally, we masked any pixel in the spectra that had low  ($<$ 50\%) fiber coverage (i.e., where less than 50\% of the light at a given wavelength from the source falls within the extracted fibers. We filled these masked pixels by interpolating the higher coverage pixels around the masked value(s).

\begin{figure}
	 \includegraphics[width=1\columnwidth]{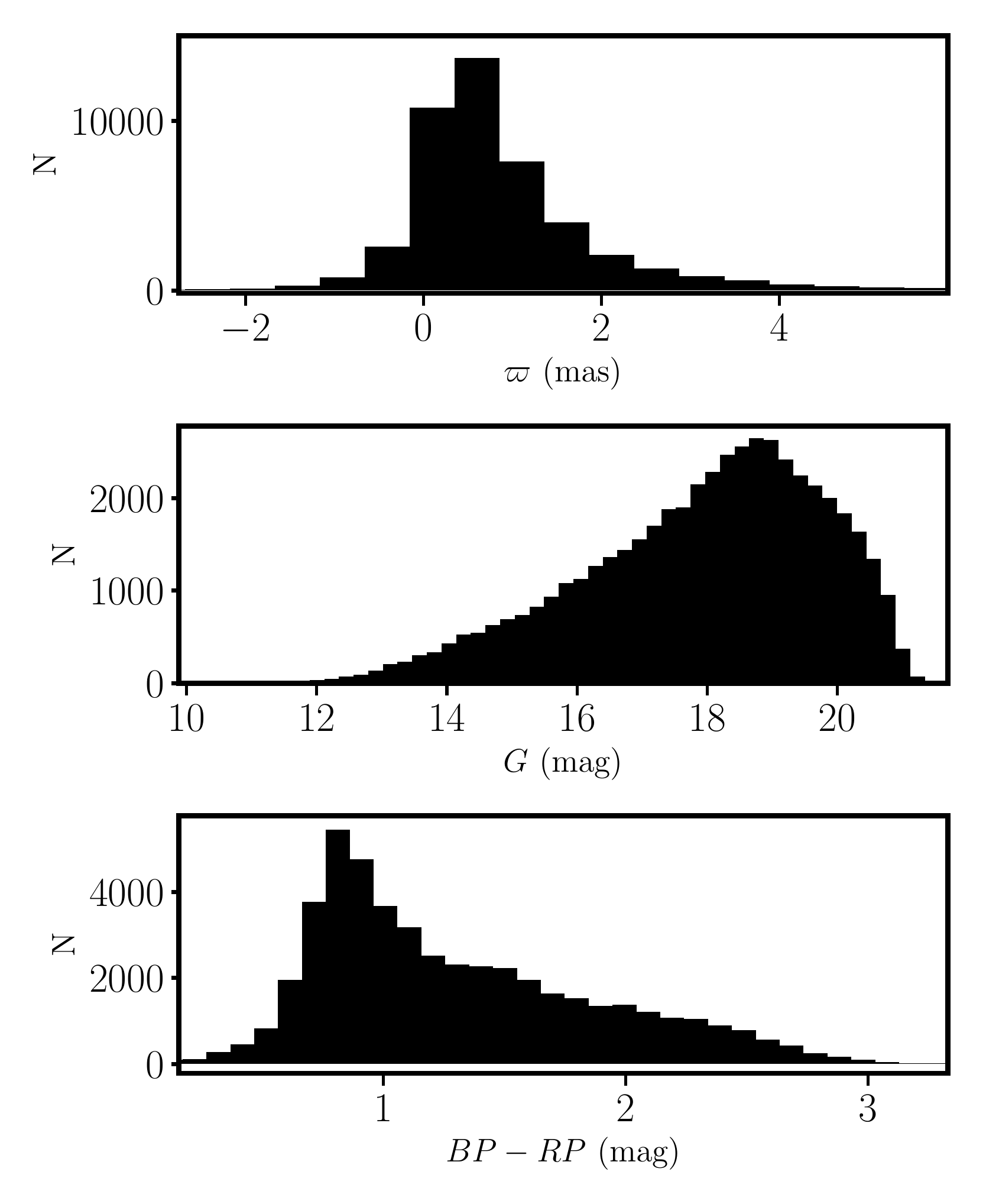}
	\caption{The distribution of parallax ($\varpi$, top panel), $G$ band magnitude (middle panel), and $BP-RP$ color (bottom panel) for the stars whose SNR $>$ 5.   } 
	\label{fig:GBPdist}
\end{figure}

We determined the signal-to-noise ratio (SNR) for each spectrum by dividing the flux by the flux uncertainty for each pixel and taking the median over the full spectral range. Figure~\ref{fig:SNR} shows the computed median SNR per pixel for the stars in the  \hetdex-\gaia\ value added catalogue as a function of their $G$ band magnitude. We generally focus on the 46,890 unique sources with SNR $>$ 5~pixel$^{-1}$ in this work. As expected, for a constant exposure time, stars brighter in \gaia\ $G$ band have significantly higher SNR. We note that there are likely issues with the brightest stars as they are affected by significant and complex scattered light.  \hetdex\ saturates around $G \sim$ 14~mag, and for sources brighter than $G \sim$ 15~mag the masking procedures can be aggressive and mistake bright spectra for cosmic rays. The observed distribution in parallax ($\varpi$), $G$, and $BP-RP$ color for stars with SNR $>$ 5~pixel$^{-1}$  can be found in Figure~\ref{fig:GBPdist}. These distributions demonstrate that the most \hetdex\ stars are farther than 1~kpc, fainter than G~$>$~18, and have colors near the turn-off.

\begin{figure*}
	 \includegraphics[width=2\columnwidth]{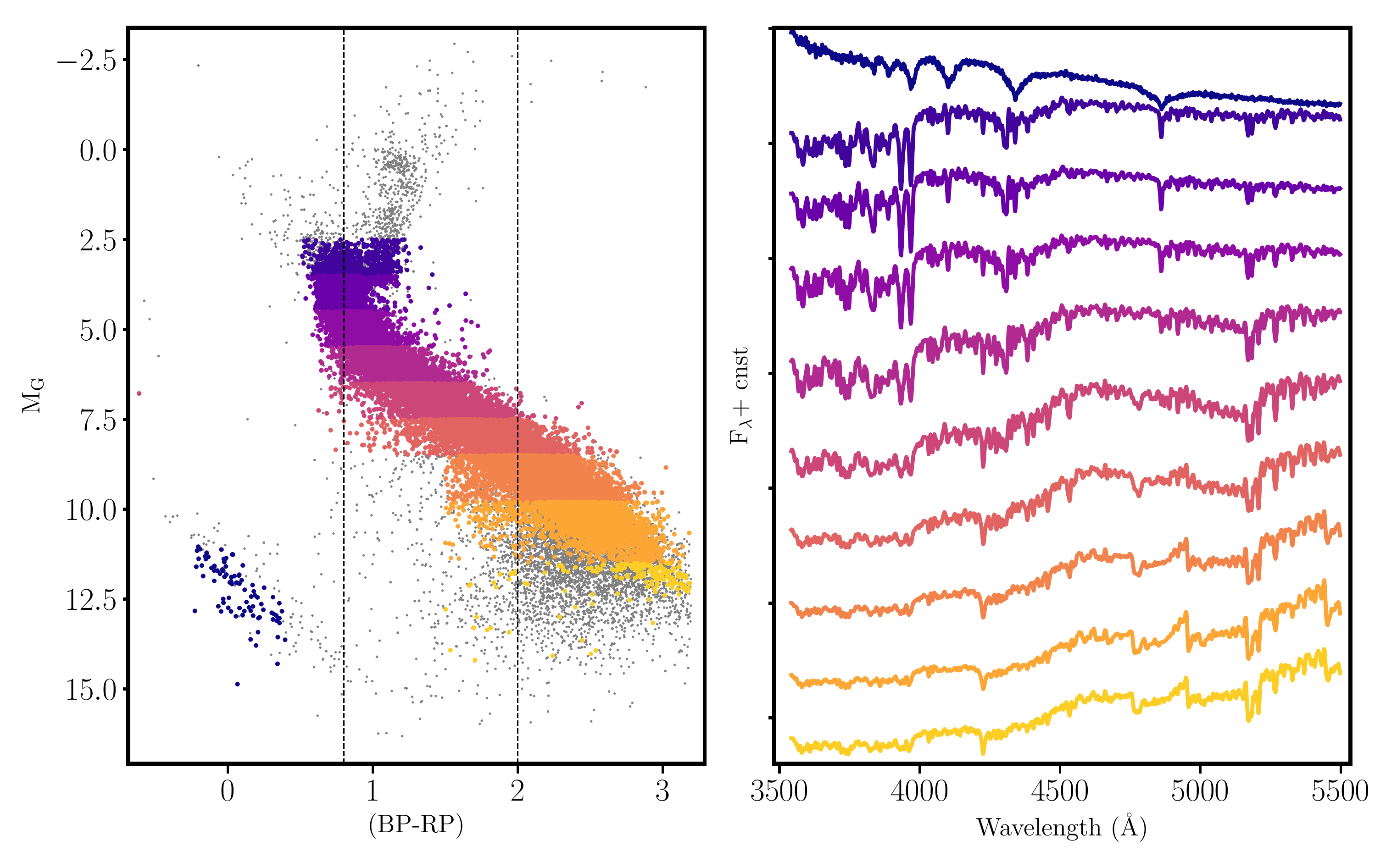}
	\caption{Left panel: The color-magnitude diagram of the \gaiahet\ value added catalogue for stars for which the parallax uncertainty is less than 30\% (gray points). The dotted black lines represent the color range $0.8 < BP-RP < 2.0$, which we focus on in this work. Right panel: The median combined spectra for stars in different regions of the color magnitude diagram with SNR $>$~5. Spectra are offset by a constant in order visualize them on the same plot. Each star represented by a colored circle in the left panel was used to generate the median stacked spectrum (with the same color) in the right panel. } 
	\label{fig:HRDspec}
\end{figure*}

To illustrate the stars uncovered in the \hetdex\ survey as well as their spectral quality, in the left panel of Figure~\ref{fig:HRDspec} we show the \gaia\ color-magnitude diagram for those stars of the \hetdex-\gaia\ value added catalogue with \gaia\ parallax uncertainty of less than 30\%. The color used for this work is the difference of the \gaia\ blue ($BP$) and red ($RP$) passbands. For this illustrative plot, we invert the parallaxes to obtain the distance and subsequently the absolute  $G$ band magnitude. However, we note that for larger uncertainties one should not invert the parallax but rather infer the distance given the parallax \citep[e.g.,][]{Bailer-Jones2015, Astraatmadja2016}. We also do not apply reddening corrections to these colors because \hetdex\ fields were chosen explicitly to have relatively low (i.e., E($B-V$) $<$ 0.05) extinction. To illustrate the typical \hetdex\ spectra for different spectral types we median-combine (or stack) stars across 10 absolute magnitude-color bins, which encapsulate the OBAFGKM-type stars (from blue to yellow, respectively) as well as white dwarfs (dark blue), that are represented by the colors on the left panel of the figure. The spectral stacks contain a hundred (white dwarfs) to a few thousand (G-type) of spectra. Stars that are denoted by gray points are those which are not used for the median stack due to low SNR (SNR $<$ 5). The right side of Figure~\ref{fig:HRDspec} shows the median-stacked spectra for white dwarfs to OBAFGKM-type stars from top to bottom. The vertical axis represents the median flux with an additive constant to offset the spectra for the purposes of visualization. Figure~\ref{fig:HRDspec} shows that \hetdex\ spectra are of high enough resolution and quality to identify key stellar features (e.g., Ca H\&K, CH G-band, H$\beta$, Mg~Triplet, TiO molecular band head, etc.).  

\subsection{Other Large Spectroscopic Surveys: LAMOST and SDSS}\label{subsec:LAMOST} 
To better understand the information content of the \hetdex\ spectra relative to other large multi-object spectroscopic surveys, we have matched several of these surveys with the \hetdex-\gaia\ value added catalogue described in Section~\ref{subsec:HETDEX}. The cross matches were preformed with the fifth data release of the \lamost\ survey, the sixteenth data release of the \apogee\ survey, and the ninth data release of the \segue\ survey. These matches were accomplished by using a $1\farcs 5$ search window in each of the surveys centered around each point source in the \hetdex-\gaia\ catalogue. We did not cross match against southern hemisphere spectroscopic surveys (e.g., \ges, \galah, \rave) because \hetdex\ was carried out in the Northern hemisphere and thus we do not expect many overlapping stars.  

The cross match with the above three surveys  have 6,364 point sources in common with the \lamost\ survey (\hetdex-\lamost\ catalogue), 5,730 point sources in common with the \segue\ survey (\hetdex-\segue\ catalogue), and 658 point sources in common with the \apogee\ survey. In Table~\ref{tab:overlap}, we summarize the depth, resolution, and overlap between the surveys discussed above. It is clear from Table~\ref{tab:overlap}, that \hetdex\ will cover a magnitude space that few stellar spectroscopic surveys cover. Given the low sample size in the cross match with the \apogee\ survey, we will not consider \hetdex-\apogee\ cross matches subsequently in this paper.

\begin{table}
\caption{The Overlap between \hetdex\ and other large surveys}
\label{tab:overlap}
\begin{tabular}{lllll}
\hline\hline
Survey & $\overline{G}$ & $\sigma G$  & Overlap N & Resolution \\
 & (mag) & (mag) & &  \\
 \hline
 \hetdex & 18.8 & 2.1 & ... & 750\\
 \segue &17.2& 1.3 &5730& 2000\\
 \lamost &15.1 & 1.9 &6364& 1800 \\
 \apogee & 13.5& 2.1 & 658 & 22500 \\

\hline\hline
\end{tabular}
\raggedright
NOTE: Summary of the overlap between \hetdex\ and other large spectroscopic surveys. Each survey considered is shown in column~1. The average ($\overline{G}$) and standard deviation ($\sigma G$) of the \gaia\ $G$ band magnitude that the survey covers can be found in column 2 and 3, respectively. The overlap between \hetdex\ and each survey and its resolution is displayed in columns~4 and 5, respectively. 
\end{table}

For a reliable comparison, we use only a subset of the \hetdex-\lamost\ and \hetdex-\segue\ catalogues. For the \hetdex-\lamost\ catalogue, we require the following: (1) the absolute value of the \lamost\ RV to be less than 600~\kms, (2)  the uncertainty in RV to be less than 10~\kms, (3) the \lamost\ SNR in the $G$ band must be larger than 5. The first three constraints were designed to ensure that the \lamost\ RV is fairly well determined. Additionally, we require (1) the absolute value of the \hetdex\ measured RV\footnote{There are $\sim$ 600 point sources in the \hetdex-\gaia\ value added catalogue that have RVs well above 600~\kms. Visual inspection of the spectra of these sources indicate that they are galaxies.} be less than 600~\kms, (2) the uncertainty in the \hetdex\ RV  be less than 50~\kms, and (3) the \hetdex\ SNR be larger than 5. These last three criteria were required to ensure that reasonable RVs were determined from the \hetdex\ spectra. For example, when the \hetdex\ measured RV uncertainty is larger than 50~\kms, which happens for a small number of stars in the \hetdex-\lamost\ and \hetdex-\segue\ datatsets, the RVs are not likely to be reliable. Together these constraints reduced the overall \hetdex-\lamost\ catalogue to 2,384 stars for RV comparisons with a median \hetdex\ SNR = 60~pixel$^{-1}$. We also impose the same set of quality criteria on the \hetdex-\segue\ catalogue, which reduced the comparison sample to 1,456~stars, with a median \hetdex\ SNR = 30~pixel$^{-1}$. These comparison samples enable robust quantification of the external precision with which we can estimate RVs using the \hetdex\ spectra (see Section~\ref{subsec:resultRV}).

\section{Methods} \label{sec:methods}
The primary aims of this paper are to present the first stellar spectra for the \hetdex\ survey, determine the RVs for the \hetdex-\gaia\ value added catalogue and find any candidate metal-poor stars hidden within the low-resolution data. As such, in the following sections, we outline the methods used to measure the RVs from \hetdex\ spectra (Section~\ref{subsec:RVs}) and identify potential metal-poor candidate stars (Section~\ref{subsec:tSNE}).

\subsection{Radial Velocities} \label{subsec:RVs}
The RVs for \hetdex\ spectra are determined using the cross-correlation technique, whereby the observed spectrum is compared to a template spectrum. To do this, we made use of the \texttt{iSpec} python package \citep{2014A&A...569A.111B}. \texttt{iSpec} cross correlates the observed spectra with a template spectrum within a specific velocity range and step. A grid of RV template spectra were generated using the TURBOSPECTRUM \citep{Alvarez1998, Plez2012} radiative transfer package along with the MARCS model atmosphere grid \citep{Gustafsson2008} and version 5 of the \ges\  linelist (Heiter et al., submitted). The same spectral synthesis configuration is also used in \cite{Hawkins2020a}, although here it is at much lower spectral resolution. These template spectra were generated with a wavelength coverage of  3500 $< \lambda <$ 5500~\AA\ with a resolution of R = 750 to approximately match the \hetdex\ spectra. The grid was generated for temperatures between 4000 $<$ \teff\ $<$ 7000~K in steps of 250~K, log gravities from 0.5 $<$ \logg\ $<$ 5.0~dex in steps of 0.5~dex, and metallicities between $-2.0 <$ \feh\ $< +0.5$~dex in steps of 0.50~dex. Given the temperature range of the grid, we restrict our further analysis to stars with 0.8~$< (BP-RP) <$~2.0. 

Once the RV template grid was constructed, each \hetdex\ spectrum was matched to one of the template spectra by minimizing the $\chi^2$. The template spectral grid was also normalized using the BMC to be consistent with the observed spectra. The cross correlation is done in velocity space with RVs between $-1000 <$ RV $<$ +1000~\kms\ and a velocity step of 1~\kms. \texttt{iSpec}\footnote{For more information consult \cite{2014A&A...569A.111B} and \url{https://www.blancocuaresma.com/s/iSpec/manual/usage/velocities}.} determines the RV by fitting a second order polynomial near the peak of the cross correlation function, and indicates if there are multiple peaks in the cross correlation, which could represent the presence of a binary companion. After the RVs are determined for each star, the barycentric correction using \texttt{iSpec} is applied to the RV\null.  As noted in the \texttt{iSpec} documentation, the uncertainties in the RV are determined using the cross correlation function and its second derivative according to \cite{Zucker2003}.

\subsection{Categorizing the \hetdex\ Stars using Machine Learning: t-SNE} \label{subsec:tSNE}
In addition to measuring the RVs for \hetdex\ stellar spectra, we are also interested in identifying any metal-poor stars in these low-resolution data, as this has not yet been done for this dataset. One way of categorizing stars on the basis of how similar their spectra appear is through the t-Distributed Stochastic Neighbor Embedding \citep[tSNE,][]{tSNE2008}. tSNE is a machine learning technique that enables the construction of a low dimensional (D), normally 2 dimensional (2-D), projection of a higher dimensional dataset while efficiently retaining much its original complexity. tSNE has been used by several other surveys to characterize their stellar spectra. For example, it has been employed by \cite{Matijevic2017} to find very metal-poor stars in the RAVE survey, by \cite{Anders2018} to explore the practicality for chemical tagging purposes with high-resolution data taken with the HARPS spectrograph, and by \cite{Jofre2017} to search for stellar twins. For the purposes of our work here, we use tSNE to explore whether the \hetdex\ spectra are of sufficient quality to find metal-poor stars, similar to \cite{Matijevic2017}, but at the lowest resolution to date and applied for the first time to this dataset. In the context of \hetdex\ spectra, tSNE will enables representation of all of the D = 1036 pixel, RV corrected low-resolution \hetdex\ spectra on a 2-D plane where stars with similar spectra are found in the same region of that space and stars that have very dissimilar spectra are separated in that space.


While a detailed description and discussion of tSNE can be found in \cite{tSNE2008}, we summarize the technique and how we apply it to the \hetdex\ spectra in this work. In order for tSNE to capture the complexity of the original high-D dataset, which in the context of \hetdex\ is the matrix of flux densities that describe all of the spectra, it must first determine the similarity between each data spectrum and all others. In this paper, we define the set of fluxes that describe a star's spectrum as \vect{f} = \{$f_{i}$\} where  \{$f_{i}$\} is the array of fluxes at a given set of $i$th wavelengths. The bolded symbols denote vectors. Additionally, the matrix of fluxes across all RV corrected, BMC normalized \hetdex\ spectra input into tSNE is defined as \vect{F}~=~\{\vect{f_k}\}\ where  \vect{f_k}\ represents the $k$th spectrum in the dataset. Throughout the text, bolded upper case symbols represent matrices. Each row of the matrix \vect{F} is an individual spectrum/star and can be thought of as a single data point in the high-D dataset that tSNE will project onto a 2-D plane. 

To describe the similarity of each spectrum against all other spectra, tSNE starts by representing each `point' (i.e., a \hetdex\ spectrum) in the high-D space as a Gaussian probability distribution. That Gaussian distribution is centered on each `point' with a dispersion of $\sigma_{i}$. The $\sigma_{i}$ is not the uncertainty in the data point but rather set by a parameter called {\it perplexity}. As noted in \cite{tSNE2008}, the {\it perplexity}, is tied to the effective number of neighbors expected in the dataset. Once tSNE represents the similarity of each spectrum against all other spectra, it will `embed' that similarity of each star with all other remaining stars in the high-D dataset through the conditional probability, $p_{j|i}$, such that :
\begin{equation}
p_{j|i} = \frac{ \exp{( - || \bm{f_i} - \bm{f_j} || ^2 /2\sigma_i^2 ) }}{ \sum_{k\neq i}  \exp{( - || \bm{f_i} - \bm{f_k} || ^2 /2\sigma_i^2  )}}
\label{eq:pji}
\end{equation}
In Equation~\ref{eq:pji}, $p_{j|i}$ is the conditional probability that the $j$-th spectrum ($ \bm{f_j}$) would select the $i$-th the spectrum ($ \bm{f_i}$) as its neighbor and thus formalizes the similarity between these two spectra. The higher this conditional probability, the more similar the data points are and vice versa. Since tSNE focuses on pairwise similarities, $p_{i|i}$ is set to 0.  As described in \cite{tSNE2008}, it is not expected that the performance of tSNE will be significantly effected by changes of perplexity between 5--50. We also confirmed that the global shape of the tSNE map of our data does not change with perplexity between these values. 

Encoded in $p_{j|i}$ (Equation~\ref{eq:pji}) is the similarity of each spectrum against all other spectra in the dataset. The next step is to embed this complex information onto a 2-D plane while minimizing the loss of information. To accomplish this, each data point in the high-D space (i.e., each \hetdex\ spectrum) is projected onto a 2-D plane, where the $i$-th spectrum in this low-D space is given by the quantity $\bm{y_i}$. As in the high-D case above, the similarity of each spectral projection in 2-D against all others, can be formalized as a conditional probability, $q_{j|i}$, such that:

\begin{equation}
q_{j|i} = \frac{ ( 1+ || \bm{y_i} - \bm{y_j} || ^2 )^{-1} }{ \sum_{k\neq i}  ( 1+  || \bm{y_i} - \bm{y_k} || ^2 )^{-1}} 
\label{eq:qji}
\end{equation}

In Equation~\ref{eq:qji}, the similarities between the 2-D spectral projection of the $j$-th ($\bm{y_j}$) and the $i$-th ($\bm{y_i}$) data points are modeled not as a Gaussian distribution but as a Student t-distribution with a single degree of freedom. We refer the reader to \cite{tSNE2008} for a detailed discussion as to why the t-distribution is conveniently chosen over a Gaussian distribution as in Equation~\ref{eq:pji}. 

The final step of tSNE is to find a projection in the 2-D space such that $q_{j|i} = p_{j|i}$, as it is in this case that the $\bm{y_j}$ and $\bm{y_i}$ are correct models of their higher-D equivalents ($\bm{f_j}$, $\bm{f_i}$). Unfortunately, this idealized condition  never occurs in practice so the approach of tSNe is to make $q_{j|i} \approx  p_{j|i}$ by minimizing the Kullback-Leibler divergence \citep[KLD,][]{Kullback59}, over all points. This KLD cost function is given by:
\begin{equation}
C = \sum_{i} \sum_{j} p_{j|i} \log{\frac{p_{j|i}}{q_{j|i}} } 
\label{eq:cost}
\end{equation}
In Equation~\ref{eq:cost}, the parameter C is the KLD cost function to be minimized, while  $q_{j|i}$ (Equation~\ref{eq:qji}) and  $p_{j|i}$  (Equation~\ref{eq:pji}) are the conditional probabilities that describe the similarity between the $j$-th and $i$-th points in the low-D and high-D spaces, respectively. 

We made use of an accelerated version of tSNE\footnote{The python implementation of tSNE we employed can be found at \url{https://github.com/lvdmaaten/bhtsne/}.} \citep{tSNE2014}, which employs the Barnes-Hut algorithm \citep{Barnes1986}. This implementation of tSNE has two tunable hyper-parameters, which include the perplexity, discussed above, and $\theta$, which ranges from 0 to 1 and sets the speed-to-accuracy of the algorithm. If $\theta$ is set to 1 the algorithm can run very fast ($\sim$few seconds) but with less accuracy. However, if  $\theta$ is set to 0, high accuracy can be achieved but at the cost of computation time. Here we choose $\theta$ = 0.1 and perplexity = 20 because our sample size is not significantly large compared to sample sizes usually put input to tSNE. The algorithm required $\sim$600~seconds to run on a single computer core for the full sample under these conditions. 

\section{Results \& Discussion: Exploring the Stars of the HETDEX Survey}\label{sec:results}
\begin{figure*}
	 \includegraphics[width=2.2\columnwidth]{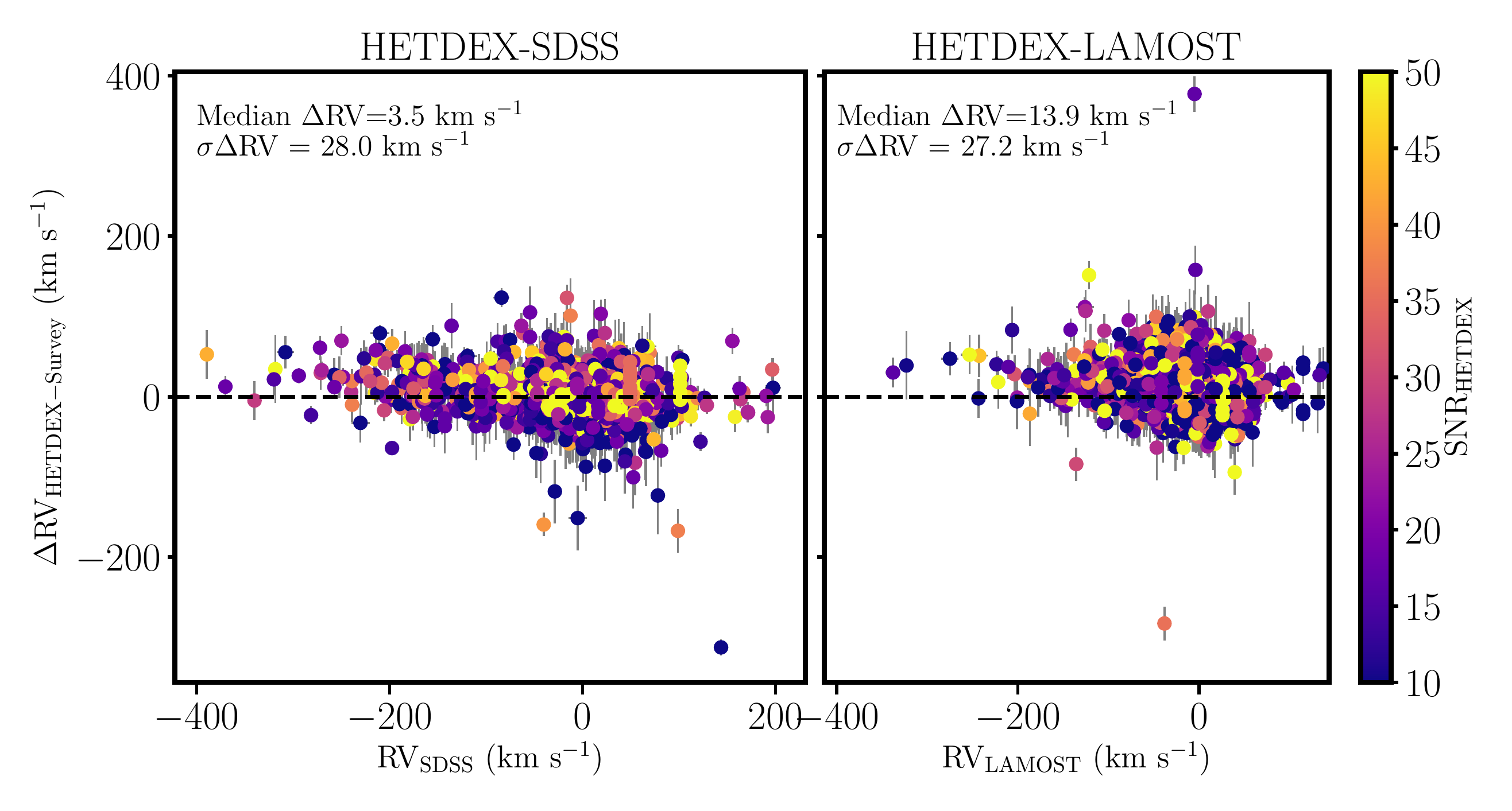}
	\caption{The difference RV determined using \hetdex\ spectra and those from the \segue\ as a function of the \segue\ RV (left panel). The same is shown on the right panel but for the \lamost\ survey. Each star is color-coded by its SNR. The dotted black line in both represent when the \hetdex\ measured RV directly matches that of measured by the external survey. The median offset and dispersion of the difference in RV, $\Delta$RV, are noted in the diagram.} 
	\label{fig:RV}
\end{figure*}
\subsection{Radial Velocities} \label{subsec:resultRV}

Our first goal was to measure \hetdex\ spectra RVs for stars with colors between 0.8 $< (BP-RP) <$ 2.0. The RVs were determined using the cross-correlation technique with a spectrum chosen from a grid of template spectra by identifying the best fitting synthetic spectrum. We refer the reader to section~\ref{subsec:RVs} for more details about the construction of the template grid, how the templates are matched to the observed spectra, and how the RVs were determined. The color range was selected in order to ensure that the \teff\ of the star was not significantly hotter or cooler than the boundaries of the RV template grid. We were able to derive RVs, their uncertainties, and heliocentric corrections for 41,980 unique stars. However, many of these spectra have SNR $<$ 5~pixel$^{-1}$ and are therefore unlikely to yield high-quality RV estimates. If we we restrict our analysis to only those spectra with SNR $>$ 5~pixel$^{-1}$, we obtain 28,343 stars. Of these 28,343 stars, $\sim$602 ($\sim$2\%) have RVs larger than 600~\kms. Visual inspection of the \hetdex\ spectra of these objects indicate that they are likely to be galaxies, so we remove any source with a measured RVs larger than 600~\kms\ from our sample.

In order to validate the quality of the RVs determined from the \hetdex\ spectra (described in Section~\ref{subsec:GaiaHETDEX}) and the cross-correlation technique (outlined in Section~\ref{subsec:RVs}), we cross matched these stars against other large spectroscopic surveys at higher resolution (i.e., \segue\ and \lamost). For the \hetdex-\segue\ cross match, which initially contained about 5,700 stars, we required that the \segue\ RV must be known with an uncertainty of less than 10~\kms\ and a \segue-derived SNR $>$ 10. We also required the \hetdex-derived RV and SNR to be less than 600~\kms. These constrained reduced the \hetdex-\segue\ comparison to 1,456 stars. For the \hetdex-\lamost\ comparison, which initially contained around 6,300 stars, we applied the same criteria as for the  \hetdex-\segue\ comparison, which reduced the  \hetdex-\lamost\ external validation sample to 2,384 stars (see section~\ref{subsec:LAMOST} for more details).

Figure~\ref{fig:RV}, presents the \hetdex\ RVs measured in this work compared with those a  measured for the same stars in the \hetdex-\segue\  (left panel) and the  \hetdex-\lamost\ (right panel) comparison samples. Each point represents an individual star and the points are color-coded by the \hetdex\ SNR\null. Figure~\ref{fig:RV} indicates that there is generally good agreement between the RVs measured by \hetdex\ and those by \segue\ with an systematic offset of +3.5~\kms\ and a dispersion in the difference in RV, $\Delta$RV, of 28.0~\kms\ (i.e., the RV measured by \hetdex\ is 3.5~\kms\ larger than the ones from \segue). There is fair agreement between the RVs measured by \hetdex\ and those by \lamost\ with a systematic offset of +13.9~\kms\ and a dispersion in the $\Delta$RV of 27.2~\kms\ (i.e., the RV measured by \hetdex\ is 13.9~\kms\ larger than the ones from \lamost). Additionally, there is a known difference of $\sim$ +9.1~\kms\ difference in the RVs measured by \lamost\ compared to those by \segue\ (whereby the \segue\ RVs are 9.1~\kms\ larger than those from \lamost) for the 66~stars that are in common between all three surveys and pass all quality criteria\footnote{This is consistent with the observed offset between the full cross match of the \lamost\ and \segue\ surveys, for example see section~3 of \url{http://dr5.lamost.org/v3/doc/release-note-v3}.}, indicating that the differences in the RV offset that we observe between the \hetdex-\lamost\ (+13.9~\kms) and \hetdex-\segue\ (3.5~\kms) validation sets can be fully accounted for by the difference in the two RV scales. There is good agreement in the external precision (measured to be $\sim$28~\kms), to which we can measure RVs relative to the two independent validation sets. We note that these results are not significantly dependent on the median  fraction of flux coverage of individual sources across fibers.

\begin{figure}
	 \includegraphics[width=1.03\columnwidth]{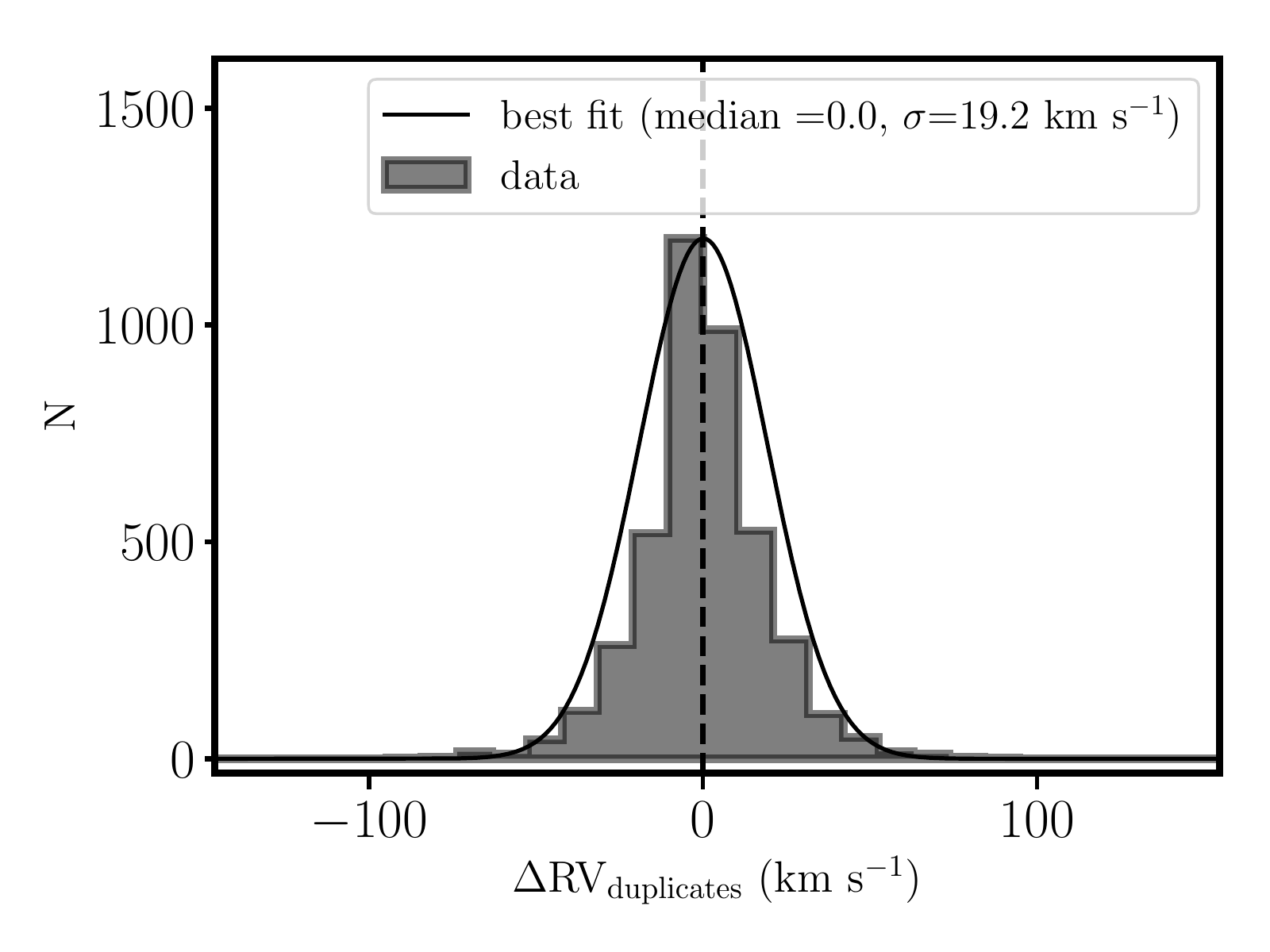}
	\caption{The distribution of the $\Delta$RV$_{\mathrm{duplicates}}$ for the $\sim$2000 stars with more than one epoch of spectra from \hetdex\null. The quantity $\Delta$RV$_{\mathrm{duplicates}}$ is the RV measured in all epoch subtracted from the median RV across all epochs for each star. The offset is consistent with zero while the dispersion, which is a measure of the internal precision, is $\sim$19~\kms (black line). } 
	\label{fig:RVdist}
\end{figure}

We can further test the measured RV precision in the \hetdex\ survey by quantifying the internal precision, i.e., by determining how consistent the \hetdex\ RVs are between stars with spectra obtained from multiple epochs. There are $\sim$2,000 unique stars which have multiple epochs of measured RVs. For each unique star with multiple spectra (and therefore multiple RV estimates), we obtain the difference in RV in each epoch and the median RV across all visits. We denote this quantity as $\Delta$RV$_{\mathrm{duplicates}}$. The distribution of $\Delta$RV$_{\mathrm{duplicates}}$ can be found in Figure~\ref{fig:RVdist}. There is an offset consistent with zero and a dispersion in $\Delta$RV$_{\mathrm{duplicates}}$ is 19.1~\kms. The dispersion in $\Delta$RV$_{\mathrm{duplicates}}$ represents a measure of the internal precision with which we can measure RVs from duplicate observations of the same (assumed to be single) stars. 

\begin{figure}
	 \includegraphics[width=1.03\columnwidth]{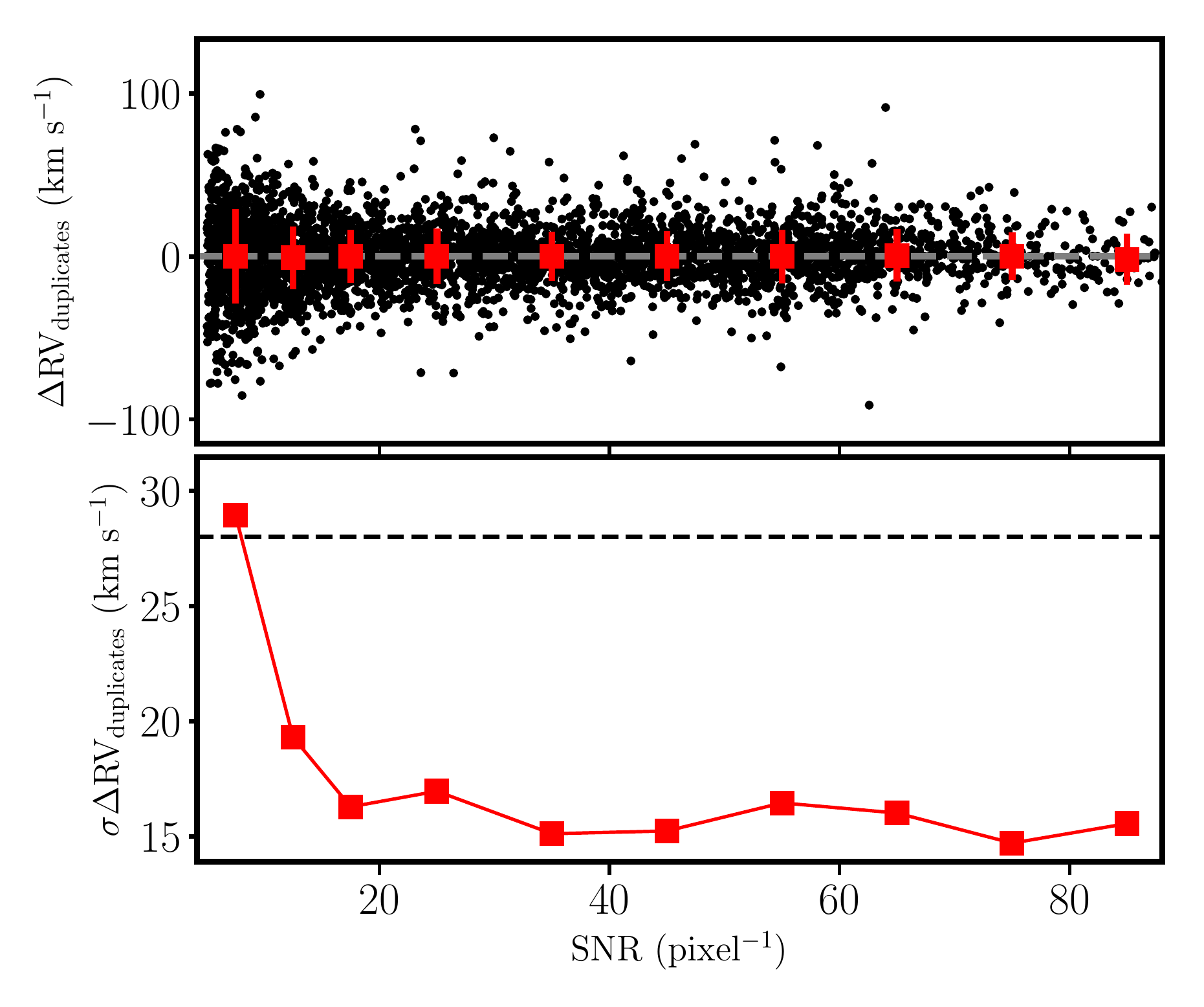}
	\caption{Top panel: The difference in the RVs measured between epochs and the median RV,$\Delta$RV$_{\mathrm{duplicates}}$,  for each star that has duplicate spectra as a function of the \hetdex\ SNR. The red squares represent the median and dispersion of $\Delta$RV$_{\mathrm{duplicates}}$ in 10 SNR bins. Bottom Panel: The dispersion in $\Delta$RV$_{\mathrm{duplicates}}$, a measure of the internal RV precision, as a function of \hetdex\ SNR. The dispersion in $\Delta$RV$_{\mathrm{duplicates}}$ is $\sim$15~\kms\ for the highest SNRs (SNR $>$ 30~pixel$^{-1}$), and rises to $\sim$29~\kms\ for the lowest SNR (SNR $\sim$ 5~pixel$^{-1}$). } 
	\label{fig:RV_SNR}
\end{figure}

Figure~\ref{fig:RV_SNR} shows how the internal precision, as measured by the dispersion in  $\Delta$RV$_{\mathrm{duplicates}}$, depends on SNR. As expected, the internal precision of RV measurements from \hetdex\ spectra degrades with decreasing SNR. The highest precision is for stars with SNR $>$ 30~pixel$^{-1}$ and the lowest is for those with SNR $<$ 10~pixel$^{-1}$. In the SNR regime that dominates the external validation sets, the internal (as measured by the duplicate spectra) and external precisions (as measure by the validation sets) are consistent. We assume that each star in the above analysis is a single star. If there were a significant fraction of unknown binaries, it would causes some of the scatter in the $\Delta$RV$_{\mathrm{duplicates}}$. 

Taken together, these results indicate that we can conservatively measure RVs in \hetdex\ at the $\sim$28~\kms\ level. We also find that the RVs measured with the \hetdex\ spectra most closely resemble those from \segue\null. However, \segue\ has an empirical correction on the order of $\sim$8~\kms\ with respect to published values from star clusters \citep[e.g.,][]{Lee2008}.  Additionally, since RVs measured with \lamost\ have a much lower zero-point correction \citep[e.g., 0.02~\kms;][]{Wang2019}, we recommend subtracting 13.5~\kms\ from the reported RVs to bring them onto the \lamost\ RV scale. 

With the RVs in hand, we correct all spectra to be in the rest-frame and move on to exploring the information content of the spectra through machine learning. 

\subsection{Finding Metal-Poor Stars in HETDEX with Machine Learning} \label{subsec:MP}

\begin{figure*}
	 \includegraphics[width=2\columnwidth]{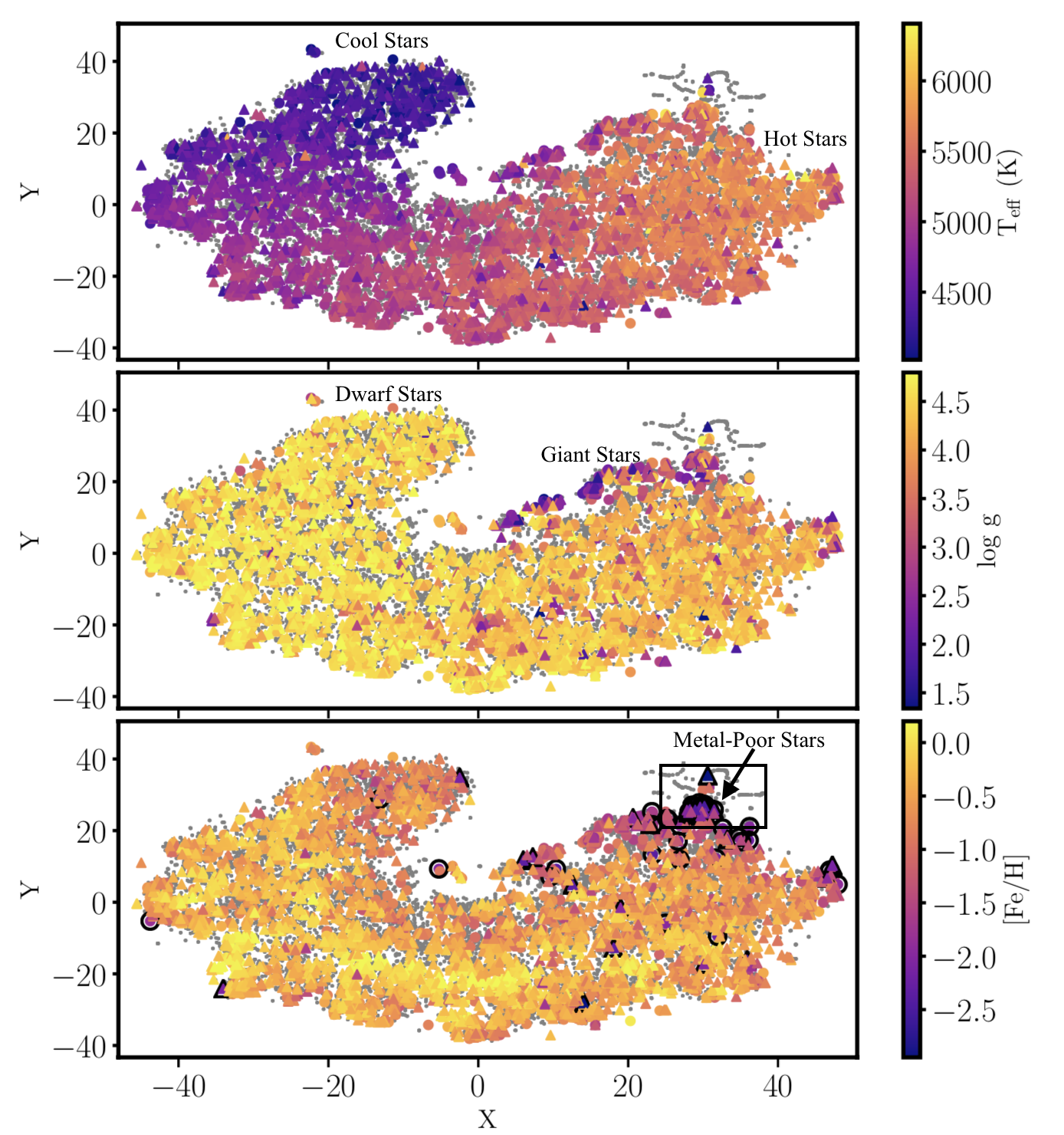}
	\caption{ Two dimensional tSNE maps (in X and Y) of the \hetdex\ sample with RV measurements (small gray points). We highlight, by color-coding, any star with a known \teff\ (top panel), \logg\ (middle panel), or \feh\ (bottom panel) from the \segue\ (colored circles) and \lamost\ (colored triangles) surveys. In the bottom panel, metal-poor stars, with \feh~$<-1.8$~dex, are highlighted with a black border around the symbol.} 
	\label{fig:tSNE}
\end{figure*}

Using the RV corrected \hetdex\ spectra, we are able to potentially identify new metal-poor stars. To achieve this, we adopt the machine learning tSNE (section~\ref{subsec:tSNE}) with the \hetdex\ spectroscopic data. We input into tSNE the BMC normalized (see Section~\ref{subsec:GaiaHETDEX}), RV corrected (see Sections~\ref{subsec:RVs} and \ref{subsec:resultRV}) spectra for stars with 0.8 $< (BP-RP) <$ 2.0\footnote{In principle, we could input all of the stars in the \hetdex-\gaia\ value added catalogue into the tSNE algorithm. This would enable us to classify each of the various types from white dwarfs to late-type stars. While we confirmed that tSNE can classify stars outside of the late-type star regime, we focused here on the FGK stars for this first paper and leave it to upcoming papers to discuss chemically peculiar stars found using this methodology.}. The tSNE algorithm requires all spectra to be sampled on the same wavelength grid with no missing pixels (due to low coverage fractions, for example). To construct this dataset, we resampled the \hetdex\ spectra onto the same wavelength grid between 3640 $< \lambda < $ 5340~\AA\ with 2~\AA\ steps. Since we initially removed fluxes where the coverage fraction is less than 0.50 (see Section~\ref{subsec:GaiaHETDEX}), we interpolated over the removed pixels during the resampling. To minimize the impact of this interpolation, we restricted ourselves to the $\sim$14,000 stars that have less than 5~pixels where the coverage fraction is less than 0.50 (i.e., we only replace as many as five~pixels via interpolation). We initially configured tSNE with perplexity of 20 and a $\theta$ = 0.10 with 1000 maximum iterations. We examined both 2-D and 3-D projections of the dataset and found little benefit to higher dimensional projection. 

Figure~\ref{fig:tSNE} shows the final projection of the high-D spectral dataset in 2-D. Each spectrum input into tSNe is represented by a closed gray circle in the figure; the final projection is color-coded by \teff\ (top panel), \logg\ (middle panel), and \feh\ (bottom panel) where those quantitites are known by either \segue\ (colored circles) or \lamost\ (colored triangles). This color-coding highlighted in this figure illustrates how the tSNE map can be used to identify regions where the spectra are similar (i.e.,  regions where the stars have similar atmospheric parameters). For example, in the final tSNE diagram, the coolest stars are in the upper corner left of center (i.e., X$\sim -10$, Y$\sim$+30) of the map (see top panel of Figure~\ref{fig:tSNE}), the central part of the tSNE map contains the majority of the spectra, and the outer edges contain more peculiar classes of stars. Critically, in the bottom panel of Figure~\ref{fig:tSNE}, demonstrates that the metal-poor stars largely clump up in the top right (i.e. X$\sim$25, Y$\sim$30) of the manifold, albeit with some spread. To help guide the eye, we place a black outline around any symbol with \feh\ $< -1.80$~dex. We also select a region where there are a high density of metal-poor stars (i.e. those with \feh\ $<$ --1.0~dex). This region contains \nummp\ candidate metal-poor stars for further investigation. 

%

\begin{figure}
	 \includegraphics[width=1\columnwidth]{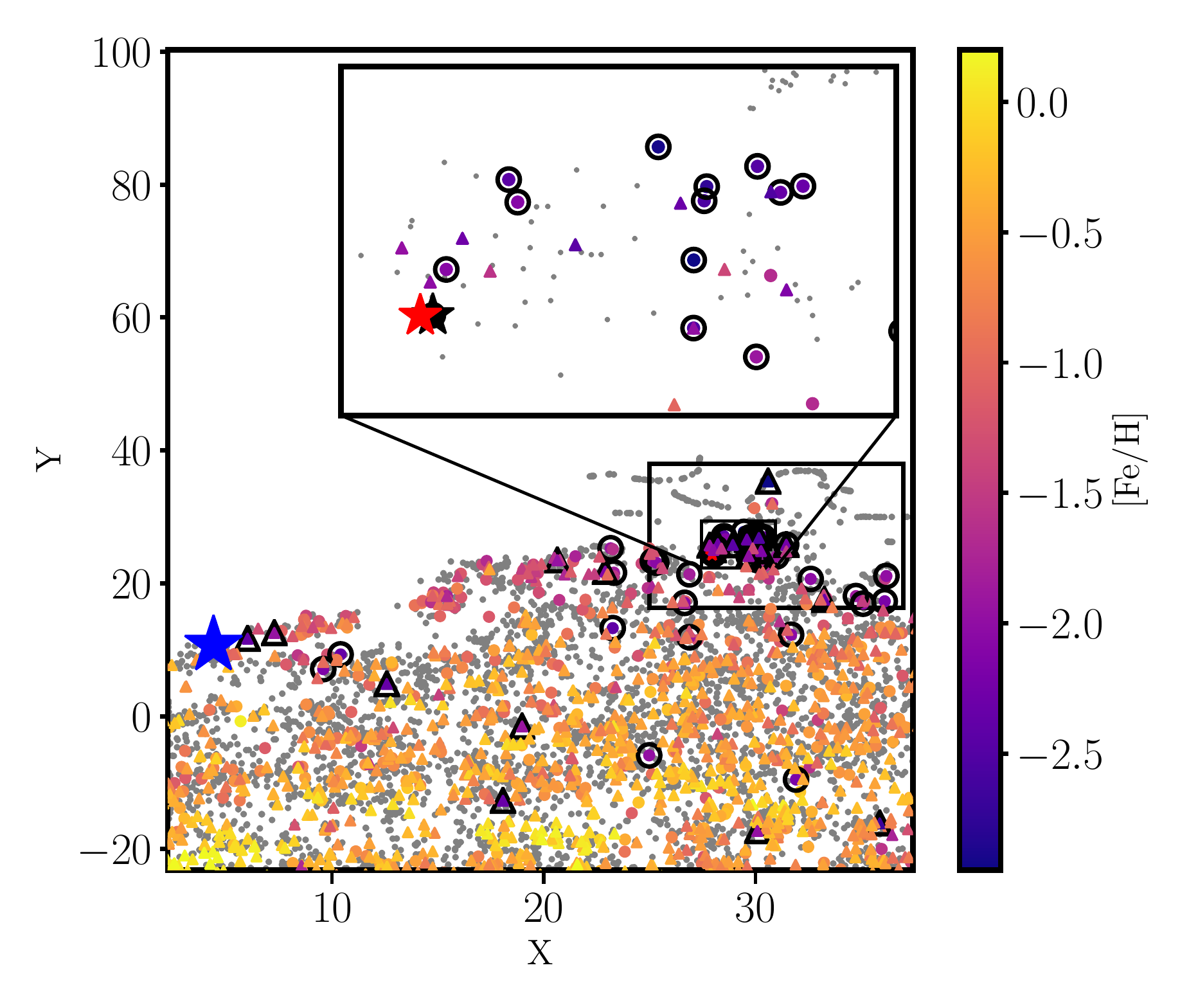}
	\caption{A zoom-in on the bottom panel of  Fig.~\ref{fig:tSNE} inside the metal-poor selection box. The inset is a further expanded view. The red star symbol is a  candidate metal-poor star  (\gaia\ DR2~1593425105811347328), which has no known metallicity from any spectroscopic survey to date. The black star symbol is a known metal-poor star (\gaia\ DR2~2537923358055035648) from \segue. For comparision, the blue star symbol is a known more metal-rich star (\gaia\ DR2~2536562574976818816) with comparable \teff\ and \logg\ values.  The black star is almost entirely behind the red star because they were selected to be direct neighbors. Nearly all stars in the region with known metallicity have \feh\ $< -1.0$~dex.} 
	\label{fig:MPselect}
\end{figure}

\begin{figure*}
	 \includegraphics[width=2\columnwidth]{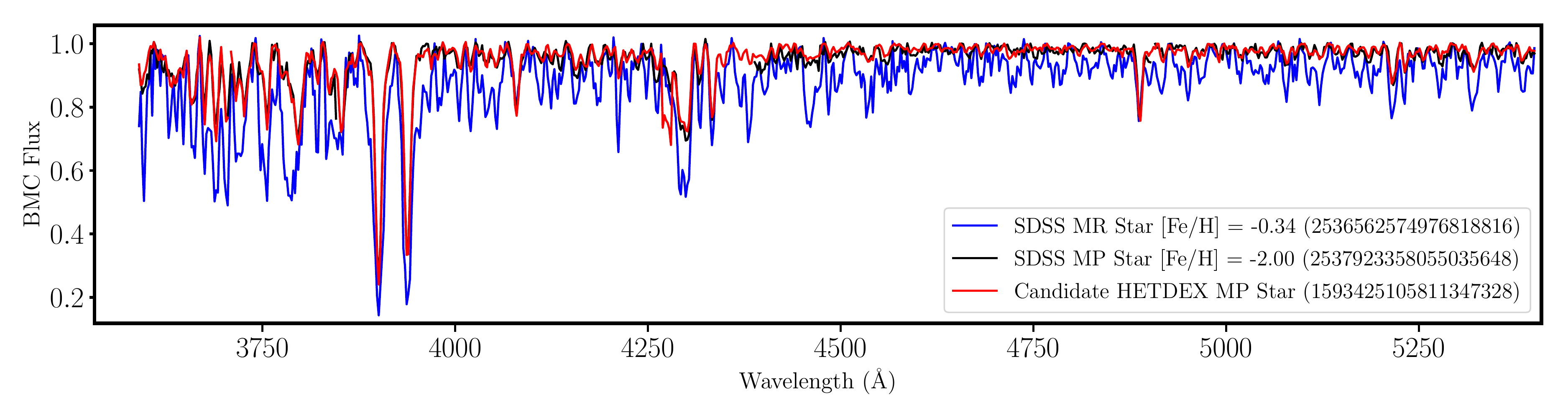}
	\caption{ The \hetdex\ spectrum of a candidate metal-poor star (red line) identified from the tSNE map in Figure~\ref{fig:tSNE}. For reference, also shown, as a black line, is a known metal-poor (\feh\ = $-2.00$) star from \segue\ that was closest to the candidate star in the tSNE diagram and a more metal-rich star (blue line), with \feh\ = $-0.34$~dex and a temperature within $\pm$200~K in \teff\ and a \logg\ within $\pm$0.20~dex of the known metal-poor star. } 
	\label{fig:MPspec}
\end{figure*}

Similar to \cite{Matijevic2017}, Figure~\ref{fig:tSNE} demonstrates that it is possible to use \hetdex\ spectra to identify stars with \feh\ $< -1.0$~dex. To further illustrate this feature, in Figure~\ref{fig:MPselect} we select a known metal-poor star  (\gaia\ DR2~2537923358055035648, black star symbol), which has a published metallicity of  \feh~=~$-2.00$~dex from the \segue\ Stellar Parameter Pipeline \citep[SSPP,][]{Lee2008, Lee2008b, AllendePrieto2008, Smolinski2011}. This figure identical to the bottom panel of Figure~\ref{fig:tSNE}, however, we have enlarged the metal-poor selection box region. We also choose the closest point in tSNE space that does not have a known metallicity (\gaia\ DR2~1593425105811347328, red star symbol). The red and black stars representing the candidate and known metal-poor stars largely lie on top of each other in Figure~\ref{fig:MPselect} Since this star is closest to known metal-poor stars in the tSNE space, it is expected that its metallicity is an excellent metal-poor candidate. For reference, the location of a known metal-rich star (\gaia\ DR2~236562574976818816), with \feh\ = $-0.34$~dex, is also shown (blue star). This star was selected such that its SSPP \teff\ was within $\pm$200~K and its SSPP \logg\ was within $\pm$0.20~dex of the known metal-poor star. 

Figure~\ref{fig:MPspec} displays the RV corrected, BMC continuum normalized \hetdex\ spectrum of the known metal-poor star (black line), the metal-poor candidate star (red line), and the metal-rich comparison star (blue line). Figure~\ref{fig:MPspec} shows that the metal-poor candidate has absorption features that are much weaker than those of the known metal-rich star, and closely resemble the metal-poor candidate star.  This indicates that the \hetdex\ spectra are of sufficient quality to find new metal-poor stars. The performance of tSNE is likely driven by the fact that \teff, \feh\, and, to a lesser degree, \logg\ information is embedded in the spectra. This could be due to the advantageous wavelength regime, which includes various stellar parameter sensitive features (e.g., Ca H\&K, Balmer lines).

We are currently obtaining observations of  the brightest of these metal-poor candidate stars with higher resolution to accurately determine their metallicities and chemical abundances. The candidate metal-poor stars discussed here represent $\sim$3\% of the total number of stars input into tSNE. Assuming the same percentage of metal-poor stars but an increase in sample size by $\sim$70\%, one would naively predict \hetdex\ to contain thousands of metal-poor stars by its competition. Importantly, since there is no preselection to objects observed in \hetdex, the final metal-poor sample from \hetdex\ will enable us explore the stellar halo of the Milky Way with a relatively simple and quantifiable selection function. 

\section{Summary } \label{sec:summary}

The Hobby-Eberly Telescope Dark Energy Experiment (\hetdex) is an ongoing blind spectroscopic survey of more than 1~million Lyman-$\alpha$ emitting galaxies. However, in its search for these galaxies, it will observe 10$^{5}$ stars. In this work, we focus on building the first catalogue of these stars and illustrate what can be done with them. We found stars in the  \hetdex\ survey by cross matching more than $\sim$200 million \hetdex\ fiber spectra with the $\sim$1.6~billion \gaia\ point sources. This enabled us find \numstar\ unique stars with \hetdex\ spectra. The spectra cover a broad range in the relatively faint magnitude regime of 10~$< G <$~21~mag. Since \hetdex\ is an blind survey\footnote{While the \hetdex\ survey design itself has no  target selection, there are several factors (e.g., magnitude limits of the survey, spectral noise, or nearby source contamination) that will may imprint biases on the underlying distribution of observed stars. Unlike for most surveys, which suffer major biases due to target selection, the stars of \hetdex\ will have a relatively simple and quantifiable selection function.}, it captures a wide range in stellar types from thousands of M-dwarfs to 100s of white dwarfs (see Figure~\ref{fig:HRDspec}). For the purposes of this work, we focused our attention on the late-type (largely FGK-type) stars by imposing a color cut of 0.8$<$ $(BP-RP) <$ 2.0. RVs were determined for these stars using cross-correlation with template spectra (Section~\ref{subsec:RVs}). The results of that analysis show that RVs derived with \hetdex\ spectra are in good agreement with those measured by the \segue\ and \lamost\ surveys. We find a median offset of 3.5 and 13.9~\kms\ with the \segue\ and \lamost\ surveys, respectively. Additionally, we demonstrate that the dispersion in the $\Delta$RV between \hetdex\ and these surveys is $\sim$30~\kms (Figures~\ref{fig:RV} and \ref{fig:RVdist}). Since \gaia\ will only be able to derive RVs for the brightest stars ($G<$14), for many of the faint stars in \hetdex\, these results will be their only RV measurement. 

We also attempted to determine whether the information content of the low-resolution spectra were enough to find metal-poor stars. To do this, we employed the tSNE machine learning algorithm which projects a high dimensional data set (i.e., the \hetdex\ spectra) onto a low dimensional manifold as a way to visualize the complexity of the data. This method has been employed in recent literature to search for metal-poor stars in moderate-resolution surveys \citep[e.g.][]{Matijevic2017}, though here we apply it at the lowest spectral resolution to date. The results of the tSNE analysis can be found in Section~\ref{subsec:MP}. Using tSNE, we uncovered \nummp\ new metal-poor candidate stars (Figures~\ref{fig:tSNE}, ~\ref{fig:MPselect} and \ref{fig:MPspec}), of which higher resolution spectroscopic followup is underway. The metal-poor stars identified here will be useful in constraining the formation of elements on the periodic table as well the early history of the Milky Way. Additionally, these results indicate that there is enough information embedded in the \hetdex\ spectra to not only derive \teff, and \logg, but \feh\ as well. 

This work represents the first attempt to extract and explore the stars observed by the \hetdex\ survey. The survey is ongoing and only $\sim$30\% complete. Thus it is expected that there will by many more stars in the final \hetdex\ sample. In an upcoming paper, we plan to validate and chemically characterize the metal-poor candidate stars discovered here with higher resolution data, use tSNE to find other types of chemically peculiar stars (e.g. carbon enhanced stars), and attempt to measure the stellar atmospheric parameters directly from \hetdex\ spectra as we now know the information content is there (e.g. see Figure~\ref{fig:tSNE}). Additionally, the results presented here provide an example of how low-resolution stellar spectra can be utilized in future surveys \citep[e.g., in the upcoming Nancy Grace Roman Telescope][]{Green2012} That combined with the blind nature of the survey will enable future stellar surveys with \hetdex\ to make a significant impact on Galactic Archaeology.

\acknowledgments

KH has been partially supported by a TDA/Scialog  (2018-2020) grant funded by the Research Corporation and a TDA/Scialog grant (2019-2021) funded by the Heising-Simons Foundation. KH acknowledges support from the National Science Foundation grant AST-1907417 and from the Wootton Center for Astrophysical Plasma Properties funded under the United States Department of Energy collaborative agreement DE-NA0003843. ES
gratefully acknowledge funding by the Emmy Noether program from the Deutsche Forschungsgemeinschaft (DFG).

The observations were obtained with the Hobby-Eberly Telescope is operated by McDonald Observatory on behalf of the University of Texas at Austin, Pennsylvania State University, Ludwig-Maximillians-Universit{\" a}t M{\" u}nchen, and Georg-August-Universit{\" a}t, G{\" o}ttingen. The HET is named in honor of its principal benefactors, William P. Hobby and Robert E. Eberly. We thank the staff at McDonald Observatory  for making this project possible.

HETDEX (including the WFU of the HET) is led by the University of Texas at Austin McDonald Observatory and Department of Astronomy with participation from the Ludwig-Maximilians- Universität München, Max-Planck-Institut für Extraterrestriche-Physik (MPE), Leibniz-Institut für Astrophysik Potsdam (AIP), Texas A\&M University, Pennsylvania State University, Institut für Astrophysik Göttingen, The University of Oxford, Max-Planck-Institut für Astrophysik (MPA), The University of Tokyo, and Missouri University of Science and Technology. In addition to Institutional support, HETDEX is funded by the National Science Foundation (grant AST- 0926815), the State of Texas, the US Air Force (AFRL FA9451-04-2- 0355), and generous support from private individuals and foundations.

The authors acknowledge the Texas Advanced Computing Center (TACC) at The University of Texas at Austin for providing computing resources that have contributed to the research results reported within this paper. URL: http://www.tacc.utexas.edu

Funding for the Sloan Digital Sky Survey IV has been provided by the Alfred P. Sloan Foundation, the U.S. Department of Energy Office of Science, and the Participating Institutions. SDSS-IV acknowledges support and resources from the Center for High-Performance Computing at the University of Utah. The SDSS web site is www.sdss.org.

SDSS-IV is managed by the Astrophysical Research Consortium for the 
Participating Institutions of the SDSS Collaboration including the 
Brazilian Participation Group, the Carnegie Institution for Science, 
Carnegie Mellon University, the Chilean Participation Group, the French Participation Group, Harvard-Smithsonian Center for Astrophysics, 
Instituto de Astrof\'isica de Canarias, The Johns Hopkins University, Kavli Institute for the Physics and Mathematics of the Universe (IPMU) / 
University of Tokyo, the Korean Participation Group, Lawrence Berkeley National Laboratory, 
Leibniz Institut f\"ur Astrophysik Potsdam (AIP),  
Max-Planck-Institut f\"ur Astronomie (MPIA Heidelberg), 
Max-Planck-Institut f\"ur Astrophysik (MPA Garching), 
Max-Planck-Institut f\"ur Extraterrestrische Physik (MPE), 
National Astronomical Observatories of China, New Mexico State University, 
New York University, University of Notre Dame, 
Observat\'ario Nacional / MCTI, The Ohio State University, 
Pennsylvania State University, Shanghai Astronomical Observatory, 
United Kingdom Participation Group,
Universidad Nacional Aut\'onoma de M\'exico, University of Arizona, 
University of Colorado Boulder, University of Oxford, University of Portsmouth, 
University of Utah, University of Virginia, University of Washington, University of Wisconsin, 
Vanderbilt University, and Yale University.

Guoshoujing Telescope (the Large Sky Area Multi-Object Fiber Spectroscopic Telescope LAMOST) is a National Major Scientific Project built by the Chinese Academy of Sciences. Funding for the project has been provided by the National Development and Reform Commission. LAMOST is operated and managed by the National Astronomical Observatories, Chinese Academy of Sciences.

This work has made use of data from the European Space Agency (ESA) mission {\it Gaia} (\url{https://www.cosmos.esa.int/gaia}), processed by the {\it Gaia} Data Processing and Analysis Consortium (DPAC, \url{https://www.cosmos.esa.int/web/gaia/dpac/consortium}). Funding for the DPAC has been provided by national institutions, in particular the institutions participating in the {\it Gaia} Multilateral Agreement. \\

%

\vspace{5mm}
\facilities{HET, Gaia, SDSS, LAMOST}


\software{astropy \citep{2013A&A...558A..33A},  
          scipy \citep{2019arXiv190710121V}, 
          t-SNE \citep{tSNE2008},
          Turbospectrum \citep{Plez2012}
          }




\bibliography{bibliography}{}
\bibliographystyle{aasjournal}



\end{document}